\documentclass{emulateapj}

\shorttitle{SLACS V}
\shortauthors{Bolton et al.\ }
 
\begin{document}
 
\title{The Sloan Lens ACS Survey. V\@.
The Full ACS Strong-Lens Sample\altaffilmark{1}}

\author{Adam~S.~Bolton\altaffilmark{2,3}}
\author{Scott~Burles\altaffilmark{4}}
\author{L\'{e}on~V.~E.~Koopmans\altaffilmark{5}}
\author{Tommaso~Treu\altaffilmark{6,7}}
\author{Rapha\"el~Gavazzi\altaffilmark{6,8}}
\author{Leonidas~A.~Moustakas\altaffilmark{9}}
\author{Randall~Wayth\altaffilmark{3}}
\author{David~J.~Schlegel\altaffilmark{10}}

\slugcomment{Accepted for publication in The Astrophysical Journal}

\altaffiltext{1}{Based on observations made with the
NASA/ESA 
\textsl{Hubble Space Telescope}, obtained
at the Space Telescope Science Institute, which is operated
by AURA, Inc., under NASA contract NAS 5-26555.
These observations are associated with programs \#10174, \#10494,
\#10587, \#10798, and \#10886.}
\altaffiltext{2}{Beatrice Watson Parrent Fellow,
Institute for Astronomy, University of Hawai`i,
2680 Woodlawn Dr., Honolulu, HI 96822 ({\tt bolton@ifa.hawaii.edu})}
\altaffiltext{3}{Harvard-Smithsonian Center for Astrophysics, 60 Garden
St., Cambridge, MA 02138 ({\tt rwayth@cfa.harvard.edu})}
\altaffiltext{4}{Department of Physics and Kavli Institute for
Astrophysics and Space Research, Massachusetts Institute of Technology,
77 Massachusetts Avenue, Cambridge, MA 02139 ({\tt burles@mit.edu})}
\altaffiltext{5}{Kapteyn
 Astronomical Institute, University of Groningen, P.O. Box 800, 9700AV
 Groningen, The Netherlands ({\tt koopmans@astro.rug.nl})}
\altaffiltext{6}{Department of Physics, University of California,
 Santa Barbara, CA 93101, USA ({\tt tt@physics.ucsb.edu})}
\altaffiltext{7}{Sloan Fellow, Packard Fellow}
\altaffiltext{8}{Institut d'Astrophysique de Paris, UMR7095 CNRS \&
Univ.\ Pierre et Marie Curie, 98bis Bvd Arago, F-75014 Paris, France
(\texttt{gavazzi@iap.fr})}
\altaffiltext{9}{Jet Propulsion Laboratory, California Institute of
Technology, 4800 Oak Grove Drive, M/S 169-327, Pasadena, CA 91109
({\tt leonidas@jpl.nasa.gov})}
\altaffiltext{10}{Physics Division, Lawrence Berkeley National Laboratory,
Berkeley, CA 94720-8160, USA ({\tt djschlegel@lbl.gov})}

\begin{abstract}
We present the definitive data for the
full sample of 131 strong gravitational
lens candidates observed with the Advanced Camera for Surveys (ACS) aboard
the \textsl{Hubble Space Telescope} by the Sloan Lens ACS (SLACS)
Survey.  All targets were selected for
higher-redshift emission lines and lower-redshift continuum
in a single Sloan Digital Sky Survey (SDSS) spectrum.
The foreground galaxies are primarily of early-type morphology,
with redshifts from $z \simeq 0.05$ to $0.5$
and velocity dispersions from $\sigma \simeq 160$\,km\,s$^{-1}$
to $400$\,km\,s$^{-1}$; the faint background emission-line galaxies have redshifts
ranging from $z \simeq 0.2$ to $1.2$.  We confirm 70 systems
showing clear evidence of multiple imaging of the background galaxy
by the foreground galaxy, as well as an additional
19 systems with probable multiple imaging.
For 63 clear lensing systems, we present
singular isothermal ellipsoid and light-traces-mass
gravitational lens models fitted to the ACS imaging data.
These strong-lensing mass measurements are supplemented
by magnitudes and effective radii measured from ACS surface-brightness
photometry and redshifts and velocity dispersions measured
from SDSS spectroscopy.  These data constitute a unique resource for
the quantitative study of the inter-relations between
mass, light, and kinematics in massive early-type galaxies.
We show that the SLACS lens sample is statistically consistent
with being drawn at random from a parent sample of
SDSS galaxies with comparable spectroscopic parameters and
effective radii, suggesting that the results
of SLACS analyses can be generalized to the massive early-type
population.
\end{abstract}

\keywords{gravitational lensing --- galaxies: elliptical --- surveys}

\section{Introduction}

Strong gravitational lensing---the multiple imaging of a distant
object by the gravity of an intervening object---provides a direct
and accurate measurement of mass in the central regions of elliptical
galaxies.  This measurement is
independent of the dynamical state of the lensing material
and nearly independent of its radial
density profile \citep[e.g.][]{kochanek_91}.
Until recently, strong lenses were relatively
rare and heterogeneously selected, a fact
which has imposed serious limitations
on their utility for statistically significant
scientific studies.  Systematic surveys employing various
observational techniques have been conducted in an
attempt to surmount this limitation.
In the radio domain, significant contributions to
the number of known galaxy-scale lenses have been
made by the survey of \citet{winn_pmn_i, winn_pmn_ii, winn_pmn_iii, winn_pmn_iv}
based on the Parkes-MIT-NRAO catalog \citep{pmn} and by the
Cosmic Lens All-Sky Survey (CLASS: \citealt{myers_class, browne_class}).
\citet{ml_rings} predicted that large numbers of
strong galaxy-galaxy lenses should be visible at
optical wavelengths.  Many such systems
have now been discovered through spectroscopic
selection of candidate objects from within the Sloan Digital Sky
Survey (SDSS; \citealt{york_sdss}) database by the Sloan Lens ACS Survey
(SLACS: \citealt{slacs1, slacs2, slacs3, slacs4, slacs6, slacs7}, hereafter
Papers~I--IV and VI--VII respectively; also see
\citealt{bolton_1402} and \citealt{bolton_mfp}, hereafter B07)
and the Optimal Line-of-Sight Survey (OLS; \citealt{ols1, ols2}).
Numerous strong galaxy-galaxy lenses and lens candidates have
also been identified through various combinations
of visual and automated inspection of large-area imaging
surveys \citep{mds_lens, fassnacht_goods,
lam_legs, cabanac_sl2s, belokurov_lens, kubo_lens, faure_cosmos}.
Finally, significant numbers of lensed quasars have been detected
through \textsl{Hubble Space Telescope} Snapshot observations of
known quasars \citep{maoz_snap, morgan_ctq327},
by high-resolution ground-based surveys of the Hamburg-ESO bright quasar catalog
\citep{he_lens_i, he_lens_ii, he_lens_iii, he_lens_iv, he_lens_v, he_lens_vi, he_lens_vii},
and by the SDSS Quasar Lens Search within the SDSS imaging database
\citep{sdssqls1, sdssqls2, sdssqls3}.
To these systematic discoveries one must also add the many serendipitously
discovered strong lenses that comprise a large fraction of the known
lens population.

Here we report the observational results of the SLACS
Survey from its initiation through the deactivation of the
\textsl{HST} Advanced Camera for Surveys (ACS) in 2007 January.
From among 131 successfully observed candidates,
we confirm a total of 70 secure strong gravitational lenses
and a further 19 possible gravitational lenses, making the
SLACS Survey the most productive strong-lens survey to date.
As a consequence of the spectroscopic selection method,
all of the SLACS lenses have known spectroscopic redshifts
for both foreground and background galaxies, giving the
SLACS sample an immediate
quantitative scientific advantage over strong-lens
candidate samples selected from imaging data.
This paper represents the definitive source for SLACS
Survey data, pending the publication of multi-color
\textsl{HST} photometry (primarily from the WFPC2 instrument)
currently being completed during Observing Cycle 16, and of a
modest number of additional lenses confirmed with WFPC2 imaging during
Cycle 15.  The organization of this work is as follows.
In \S\ref{sample} we describe the candidate selection and
\textsl{HST} observing strategy.  Section~\ref{datared}
presents our data-reduction procedures.  Section~\ref{photomet}
describes the photometric modeling methods that we apply
to the images of the foreground galaxies.  We employ both
elliptical radial B-spline models
(to obtain detailed light profiles and to generate
residual images for strong lens modeling) and
elliptical \citet{devauc} models (to measure global magnitudes and
structural parameters).  The details of our strong gravitational
lens analysis are presented in \S\ref{lensan}.
Our lens classification procedure and
an overview of the resulting lens sample is presented
in \S\ref{lenshyp}.  Section~\ref{lensmod} describes our
strong-lens mass modeling procedure as applied to
63 of the secure strong lens systems, yielding the
aperture-mass measurements that enable the
scientific applications of the sample.
In \S\ref{mcomp} we compare our measurements
with quantities obtained through other methods, as a
cross check and in order
to make realistic estimates of our measurement errors.
Section~\ref{compsample} examines the
representativeness of the SLACS lenses among early-type galaxies in general.
We summarize and offer some concluding remarks in \S\ref{sumcon}.
Appendix~\ref{dataapp} provides complete data tables and image figures,
as well as comments on
the 7 secure lenses that do not admit simple lens-modeling analysis.

Throughout this work, we assume a general-relativistic
Friedmann-Robertson-Walker (FRW) cosmology with matter-density parameter
$\Omega_M = 0.3$, vacuum energy-density parameter
$\Omega_{\Lambda} = 0.7$, and Hubble parameter
$H_0 = 70$\,km\,s$^{-1}$\,Mpc$^{-1}$.  Magnitudes are
quoted in the AB system.

\section{Sample selection and observations}
\label{sample}

The gravitational lenses presented in this work were all selected
from the spectroscopic database of the SDSS based on the presence of
absorption-dominated galaxy continuum at one redshift and nebular
emission lines (Balmer series, [O\textsc{ii}]\,3727,
or [O\textsc{iii}]\,5007) at another, higher redshift.  The spectroscopic
lens survey technique was first envisioned by~\citet{warren_0047_96}
and \citet{hewett_00}
following the serendipitous discovery of the gravitational lens
0047$-$2808 through the presence of high-redshift Lyman-$\alpha$
emission in the spectrum of the targeted lower redshift
elliptical galaxy.
Further details of our particular approach are provided
in \citet{bolton_speclens} and Paper~I\@.
The SLACS Survey includes candidates from the SDSS MAIN galaxy sample
\citep{strauss_main} in
addition to candidates from the SDSS
luminous red galaxy (LRG) sample \citep{eisenstein_lrg}.
Most candidates were selected on the basis of multiple emission lines,
though several lens candidates were observed under \textsl{HST} program \#10886 on the
basis of secure [O\textsc{ii}]\,3727 line detections alone.
By virtue of this spectroscopic selection method,
all SLACS lenses and lens candidates have secure foreground
(``lens'') and background (``source'') redshifts from
the outset.  Accurate redshifts such as these
are essential to all quantitative
scientific applications of strong lensing.

From among the set of spectroscopically identified candidates,
target lists for follow-up \textsl{HST} imaging observations were
created based on a number of competing considerations: (1) maximal
nominal lensing cross sections, as determined from foreground and
background redshifts and SDSS velocity dispersions using a singular
isothermal sphere model; (2) a reasonably
uniform distribution in lens redshifts and velocity dispersions, within
the limits of feasibility; and (3) the significance of the spectroscopic
detection of background emission lines.

The selected candidates were observed under three discovery
programs: \#10174 (Cycle 13, PI: L. Koopmans), \#10587
(Cycle 14, PI: A. Bolton), and \#10886 (Cycle 15, PI: A. Bolton).
Program \#10174 was executed as a Snapshot program, with two 420-s
exposures per visit: one through the F435W filter and one through
the F814W filter.  Program \#10587 was originally implemented identically
to \#10174, but the F435W exposures were canceled early in the
observing cycle, since the advent of 2-gyro \textsl{HST} guiding
had significantly reduced Snapshot program execution rates relative
to previous cycles.  This reduction in Snapshot execution rates
somewhat compromised the specific goal of program \#10587 to obtain
a greater number of lower-mass gravitational lens galaxies,
which have a lower confirmation rate by virtue
of their smaller lensing cross section.
Nevertheless, as seen in B07 and Paper~VII, the resulting combined
SLACS lens sample has sufficient leverage in mass to define
mass-dynamical and mass-luminosity scaling relations for the
luminous early-type galaxy population.
Program \#10886 was executed as a General Observer (GO) program,
with one orbit per target through the F814W filter, split among
four closely dithered pointings.  New lenses confirmed by these
discovery programs were subsequently scheduled for observation
with full orbits and through complementary filters under programs \#10494, \#10798,
and \#11202 (Cycles 14, 15, and 16, respectively; PI: L. Koopmans).
All programs used the Wide-Field Channel (WFC) of the ACS until
the untimely demise of that camera in 2007 January prompted a transfer
of the program to WFPC2.

The work presented here is based on the full SLACS \textsl{HST}-ACS dataset,
and includes data from all SLACS programs except \#11202,
which is carried out entirely with the WFPC2.  The analysis
in the current paper makes exclusive use of the
F814W ($I$-band) data, since all ACS targets were observed at
least once through this filter.  Multi-color coverage
of the SLACS lens sample is currently being obtained under
program \#11202; multi-band results based on ACS, WFPC2, and NICMOS data will
be published following the completion of \textsl{HST}
Observing Cycle 16.

\section{Data reduction}
\label{datared}

All ACS frames were downloaded from the online archive at
the Space Telescope Science Institute on 2007 April 03,
having been processed by version 4.6.1 of the \texttt{CALACS}
calibration software.
The following steps were applied to all frames, after the generation
of a catalog file associating multiple exposure, filters, and visits
to the same unique target with one another:

\begin{enumerate}
\item From the ``FLT'' file, extract the central
1500$\times$1500 pixel
(roughly $75\arcsec \times 75 \arcsec$) section of the ACS WF1
aperture, in which the targets were centered.
\item Subtract the sky level as determined by the \texttt{MULTIDRIZZLE}
software and recorded in the \texttt{MDRIZSKY} header parameter.
\item Identify and mask significantly negative ``cold pixels'' in the cutout,
then process the cutout with the L.A. Cosmic software
(\citealt{lacosmic}, as
implemented in IDL) in order to identify and mask cosmic rays (CRs).
\item Tabulate manually the approximate pixel location of the target
galaxy in each exposure.  For multi-exposure visits, obtain the
approximate shift between exposures through image cross correlation.
\item Use the distortion information in the fits headers to generate
tangent-plane RA and Dec coordinate images relative to a fixed reference pixel.
\item Find the centroid of the target galaxy in each frame
by fitting an elliptical Moffat profile as a function
of RA and Dec (without point-spread
function convolution) to the image using
the \texttt{MPFIT2DPEAK} non-linear fitting routine in IDL.
\item Rectify the individual frames onto a uniform $0.05\arcsec$ grid
(centered on the RA and Dec centroid from the previous step)
via bilinear interpolation within the images as dictated by the distortion
solution.  Also rectify, with identical sampling,
an appropriate model point-spread function
(PSF) as determined by the Tiny Tim software \citep{tinytim}
using an input spectral energy distribution equal to the
median of all normalized SDSS spectra of SLACS targets.
\item Divide the counts and count-errors of each
frame by the exposure time to convert to counts per second.
\item For sets of multiple dithered exposures, combine all exposures
into a single stacked exposure, with an additional CR-rejection step.
Similarly, combine the PSF samplings corresponding to the
individual exposures.
\item Visually classify all targets for multiplicity and
morphology.  Systems with two or more foreground galaxies of
comparable luminosity are classified as ``multiple'', while systems
with only a single dominant foreground galaxy are classified
as ``single''.  Morphological classification is made by
a consensus of the authors through the inspection
of F814W ACS data alone, and is limited to the
categories of ``early-type'' (elliptical and S0),
``late-type'' (Sa and later spirals), and ``unclassified''
(generally ambiguous between S0 and Sa).
\end{enumerate}

We adopt this recipe in preference to the
\texttt{MULTIDRIZZLE} reduction package because the
``drizzle'' re-sampling algorithm \citep{drizzle}
is not well suited to
single-exposure Snapshot data.  By using the above
reduction procedure for both
Snapshot and dithered multi-exposure imaging data,
we guarantee that our analysis is as uniform as possible.

\section{Photometric measurement}
\label{photomet}

This section describes the details of our ACS F814W
surface photometry.  This photometric modeling pertains
exclusively to the bright, foreground galaxy in each candidate
lens system: i.e., the ``lens'' in the case of a bona fide strong lens system.
These photometric models serve to characterize
the brightnesses, sizes, and shapes of the foreground
galaxies, as well as to generate model-subtracted residual images
of the background galaxies suitable for the strong-lensing
classification and modeling described in \S\ref{lensmod}.
Depending upon the particular application, we use either
radial B-spline models or \citet{devauc} models.
Direct F814W images of all ACS targets are shown in Figure~\ref{allimage}
(in Appendix~\ref{dataapp}).

\subsection{Radial B-spline analysis}
\label{bsphot}

SLACS provides a sample of bright lensing galaxies with
relatively faint lensed galaxies in the background.
While this is a benefit to the study of the lens galaxies themselves,
it presents a challenge for strong-lens mass models that must
be fitted to those faint lensed features.
We address this challenge with the radial B-spline galaxy image
modeling technique, introduced in Paper~I\@.
Radial B-splines provide a generalized basis for modeling the radial luminosity
profile of early-type galaxies, including low-order angular effects
through the inclusion of multipole terms.
By virtue of their significant freedom, the radial B-spline models
are able to produce very cleanly subtracted residual images of the
(often lensed) background galaxies; by contrast, the best-fit
de Vaucouleurs or \citet{sersic_law} models in many cases
leave systematic residuals at count levels comparable to those of the
relatively faint strongly lensed features.

In this work, we use radial B-spline models not only to
generate residual images, but also as the basis for
aperture photometry and light-traces-mass lens models
(see \S\ref{lensmod} below).
Motivated by this goal, we implement the modeling in
a somewhat different manner than in Paper~I,
incorporating an overall isophotal ellipticity
and solving for PSF-deconvolved models.
Specifically, we define a generalized elliptical radial coordinate,
\begin{equation}
R_{\mathrm{ell}} = \sqrt{q x^2 + y^2 / q}~,
\end{equation}
where the $(x,y)$ coordinate system has the lens galaxy center at its
origin and is aligned with the principal axes of the galaxy image.
The lens-galaxy light profile is then modeled as a B-spline function of
$R_{\mathrm{ell}}$ as described in Paper~I, with
the lens-galaxy isophotal axis ratio $q$, the position
of the lens center $(x_{\mathrm{c}},y_{\mathrm{c}})$, and the major-axis
position angle of the galaxy image (measured E from N)
as non-linear model parameters
in addition to the linear B-spline coefficient amplitudes.
For a given trial choice of
the non-linear parameters, basis images corresponding to the B-spline
coefficients are generated and convolved with the appropriate PSF,
and the linear combination
of these basis images that best fits the data is computed.

We perform the B-spline model fits to the sky-subtracted
imaging data over a $14 \arcsec \times 14 \arcsec$ region
centered on the target lens candidate galaxies.
The box size is chosen primarily to extend well beyond the
scale of all lensed features and half-light radii.
Before fitting, we manually generate masks for
stars, neighboring galaxies, and possible lensed features
so as to exclude those pixels from the fits.
The initial B-spline modeling includes no higher-order multipole
terms, and solves for the non-linear parameters
by minimizing the $\chi^2$ statistic using
the IDL \texttt{MPFIT} implementation of the
Levenberg-Marquardt algorithm \citep{more_wright}.
The residual images produced by subtracting these initial models
are then examined, and the masks are manually grown to
exclude features not flagged in the original images.
A second round of B-spline models is then computed by fixing the
non-linear parameters and allowing for the following
combinations of multipole terms in the fit: none, quadrupole,
quadrupole$+$octopole, dipole, dipole$+$quadrupole.  The inclusion
of these terms allows the model to fit the effects of diskiness/boxiness,
isophotal twist, variable ellipticity with radius, and an imperfect
PSF model.  We inspect
the residual images generated by subtracting these model fits and
select a particular multipole
combination.  Models are preferred in the order given
in the preceding list, with later models being adopted only if they
provide visibly significant improvement over earlier models.
The inclusion of dipole terms is necessary for some systems in order
to model slight asymmetry in the galaxy image.  A small number
of systems (mostly edge-on S0s) require multipole orders beyond
the simple list; those systems are handled separately, with additional
multipole orders added until the residual images are satisfactory
for strong-lensing analysis.  This special handling is only done for
systems whose direct images show possible evidence of strong
lensing (see Figures~\ref{allimage} and~\ref{prettypics} in
Appendix~\ref{dataapp}).

\subsection{De Vaucouleurs analysis}
\label{devphot}

To compute standardized model magnitudes, effective radii $R_e$,
and projected axis ratios of the SLACS targets,
we fit the images
with two-dimensional ellipsoidal de Vaucouleurs luminosity profiles.
These fits are performed over a $51 \arcsec \times 51 \arcsec$ square region
centered on the target galaxies (approximately half the narrower dimension of the
WF1 CCD aperture in which the targets were rougly centered).
The manually created masks from
the B-spline stage are applied in the central regions; stars
and neighboring galaxies outside the manually masked area
are masked from the de Vaucouleurs fit with a single-step
``clipping'' of pixels that deviate by more than 4 sigma higher than the model.
The fits are performed using the \texttt{MPFIT2DFUN} procedure in IDL,
and include convolution with the appropriate rectified and stacked
Tiny Tim PSF.  The initial optimization
is done by sampling the model at one point per
data pixel; a final optimization
is done with $5 \times 5$ sub-sampling per pixel.
Model magnitudes are computed from the full (not truncated)
analytic integral of the best-fit de~Vaucouleurs model.
Effective radii are quoted at the intermediate axis:
i.e., the geometric mean of the major and minor axes of the
elliptical isophotal contour enclosing one-half the model flux.

To test for bias in the de~Vaucouleurs model-based
magnitude measurements, we compare to aperture fluxes
evaluated using the more general
B-spline luminosity-profile models of \S\ref{bsphot}.
We consider an aperture defined by twice the de~Vaucouleurs
effective radius, which in the de~Vaucouleurs case
encloses 69\% of the total model flux.
We exclude ten galaxies whose effective radius exceeds the
range modeled by the B-spline method above.
The mean fractional difference (B-spline minus de~Vaucouleurs)
in aperture flux values
across the sample is 1.0\%, with an RMS difference of 2.3\%.
Thus we see that the de~Vaucouleurs magnitudes are in good agreement
with magnitudes determined through less parametric methods.

In order to obtain rest-frame photometric quantities, we apply several
corrections to the observed $I$-band magnitudes.  We apply
corrections for Galactic dust extinction using the values of \citet*{sfd_dust}.
We also apply $k$-corrections to transform observed
$I$-band magnitudes to rest-frame $V$-band magnitudes:
these two passbands are very well matched
for the higher redshift SLACS lenses, and reasonably
close in wavelength for the lower redshift lenses.
Since multi-band observations are not available for the
full target sample, and since the SDSS colors will in general
be affected by contributions from the background galaxies, we apply
a single redshift-dependent $k$-correction based upon
a single-burst synthetic stellar population \citep{bc_2003},
as described in \citet{treu_2001_ii}.  These same $k$-corrections
were used in the analysis of Paper~IV, and should be well suited
to the old stellar populations fund in the SLACS lenses
(see Paper~II).  We expect these $k$-corrections to
be accurate to better than 0.05\,mag \citep{macarthur_07}.
Forthcoming multi-band
\textsl{HST} photometry for the full SLACS lens sample
will permit measurement of lens-galaxy colors separately from
those of the background galaxies, thus enabling the most
accurate $k$-corrections.
The $k$ corrections applied in
B07 included a computational
error that has been corrected in the current analysis
(and that does not alter the conclusions of that work, as can be
seen in Paper~VII)\@.
We derive corrections to absolute luminosity using
the adopted $(\Omega_M, \Omega_{\Lambda}, h) = (0.3,0.7,0.7)$ FRW cosmology.
Finally, we correct for luminosity evolution in the sample assuming a rate of
$d \log L_V / dz = 0.4$
\citep{kelson_2000_iii, treu_2001_iii, moran_2005_iii},
derived from the evolution of the fundamental plane
relationship \citep[FP][]{dr_fp, dd_fp}.
Ideally we would like to constrain this evolution rate
directly within the SLACS sample, but the sample probes
systematically more massive and luminous galaxies at higher redshift,
and thus evolutionary trends are significantly covariant with
mass/luminosity trends (see Figure~\ref{distribs} in \S\ref{lenshyp} below).
The evolution correction that we apply here is the same
as was adopted in Paper~IV, though here we
correct luminosities to $z = 0$ rather than $z = 0.2$.
In either case, the RMS variation about the sample
mean luminosity correction is on the order of a few hundredths dex,
since the SLACS sample does not span an especially wide range in redshift.
The measured photometric parameters for the full SLACS target sample
are presented in Table~\ref{obstable} (in Appendix~\ref{dataapp}),
along with SDSS names/coordinates, redshifts, and velocity dispersions.

\section{Strong lensing analysis}
\label{lensan}

This section presents the details of our strong-lensing
analysis.  The first evidence in support
of the strong-lensing hypothesis is the presence
of two distinct galaxy redshifts within the same SDSS
spectrum, covering a 3$\arcsec$ diameter spatial region,
which forms the basis of our \textsl{HST}-ACS target selection.
Further evidence is provided by the appearance of features
characteristic of strong lensing in our high-resolution
\textsl{HST} follow-up imaging, by successful quantitative
strong-lensing models of those features, and in some cases
by spatially resolve spectroscopy of the background-redshift
emission-line flux.

\subsection{Classification and sample overview}
\label{lenshyp}

The classification of observed candidates into lenses
and non-lenses is made by visual examination of the direct
and B-spline model-subtracted residual images in
all available \textsl{HST}-ACS bands, based on the
appearance of arcs, rings, and multiple images centered on the
position of the foreground galaxy.  Initially, this classification
is made independently by three different subsets of
the authors (ASB, RG, and LVEK $+$ TT).  Out of the systems selected
as definite lenses by any one individual initial judgment, the
percentage of unanimously agreed-upon definite lenses ranges
from 77\% to 87\%.  Subsequently, all systems are inspected
simultaneously by a single group of authors (ASB $+$ LVEK $+$ TT $+$ LAM),
and a consensus classification into definite lenses
(``grade A''), possible lenses (``grade B''),
and non-lenses or systems of unknown status (``grade X'')
is decided, additionally
taking into account integral-field spectroscopic
evidence where available (see below).  In the case of grade-A
systems, the
ACS direct and residual images show clear evidence of multiple imaging
of a background galaxy consistent with general strong-lensing
geometries.  For grade B systems, the ACS data show
evidence of probable multiple imaging, but have either a
signal-to-noise ratio (SNR) too low for
reliable lens modeling and definitive conclusion,
or some degree of ambiguity in the identification of
lensed features.  We anticipate that the majority
of the grade-B systems will be promoted to grade-A
upon the completion of deeper imaging in multiple
bands.  Grade X is
a catch-all classification that includes systems where the background galaxy
is only singly imaged (i.e., positioned at large impact parameter
relative to the foreground galaxy) and systems where the likely source
of background-redshift line emission is either undetected
or very weakly detected in the
ACS imaging.  In principle, grade-X systems
with a background galaxy at large impact parameter
are also a matter of insufficient
SNR, since at arbitrary imaging depth
some part of any background galaxy
may be seen to be strongly lensed.  However, practical confirmation
and measurement seems out of reach for these systems.
The consensus classifications of all ACS targets
are given in Table~\ref{obstable} in Appendix~\ref{dataapp}.
Out of a total of 131 successfully observed targets,
we confirm a total of 70 grade-A lenses, 19 grade-B lenses,
and 42 non-lenses (grade X).
The numerical breakdown of lenses confirmed in each of the three discovery
programs (\#10174, \#10587, and \#10886) is presented in Table~\ref{tot_table}.

\begin{figure*}
\plotone{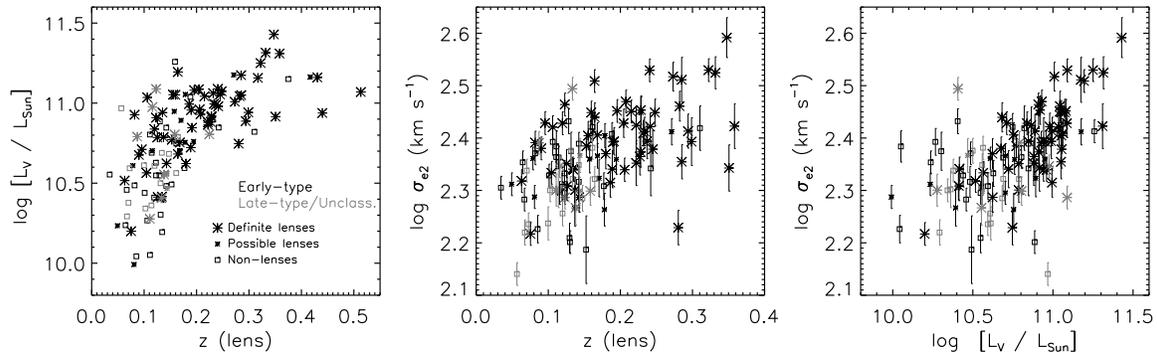}
\caption{\label{distribs}
Joint distribution of ACS targets in redshift,
luminosity, and velocity dispersion.}
\end{figure*}

Figure~\ref{distribs} shows the distribution of SLACS targets
and confirmed lenses in redshift, velocity dispersion, and luminosity.
One can see the significant covariance between magnitude
and redshift---fundamentally a consequence of the SDSS spectroscopic
target selection---that prevents us from using the SLACS lens sample to
track the evolution of a single population across redshift.
We must rather assume a rate of luminosity evolution as we have done
here, or alternatively assume that the sample evolves onto
the locally observed FP relation at redshift
$z=0$.  This latter approach will be feasible
once multi-band photometry of the SLACS lens sample is complete.

\begin{table}[t]
\begin{center}
\caption{\label{tot_table} Summary of SLACS lens discovery programs (ACS only)}
\begin{tabular}{cccc}
\hline \hline
Program & Grade-A & Grade-B & Grade-X \\
Number  & Lenses  & Lenses &  Systems \\
\hline
10174 &  26 &   5 &   8 \\
10587 &  16 &  10 &  28 \\
10886 &  28 &   4 &   6 \\
Total &  70 &  19 &  42 \\
\hline
\end{tabular}
\end{center}
\end{table}

\begin{figure}
\plotone{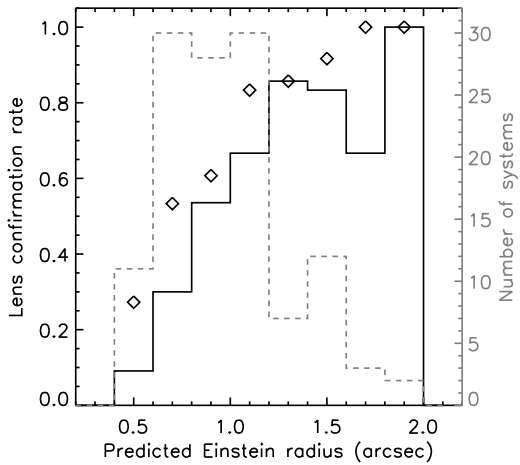}
\caption{\label{ratefig}
SLACS lens confirmation rate as a function of predicted Einstein radius
$\theta_{\mathrm{E}}$.
Values for $\theta_{\mathrm{E}}$ are computed from foreground and background
galaxy redshifts and velocity dispersions, all measured from SDSS
spectroscopy, in combination with a singular isothermal sphere galaxy
model.  Only systems with a median SNR of 10 or more
per 69-km\,s$^{-1}$ pixel over the rest-frame range 4100\AA\ to
6800\AA\ are considered here, so as to ensure well-measured
velocity dispersions.  Solid black line shows
lens confirmation rate (left-hand ordinate)
for grade ``A'' lenses, while black
diamonds indicate the confirmation rate for
grade ``A'' and ``B'' lenses combined.
Dashed gray line shows total number of
targeted systems in each bin (right-hand ordinate).}
\end{figure}

One of the powerful aspects of the selection of SLACS targets from
within the SDSS spectroscopic database is the ability
to estimate the angular lensing Einstein radius $b$
(and hence the strong-lensing cross section) of candidates before
follow-up observation.  This is possible
through the combination of foreground and
background spectroscopic redshifts with measured SDSS
velocity dispersions and a simple singular isothermal sphere
model as per Equation~\ref{b_sie}.
The conversion from lensing \textit{cross section}
to lensing \textit{probability}
requires a knowledge of the distribution of background galaxies in
size and luminosity, as well as an accounting for
the footprint of the SDSS fiber projected back into
the un-lensed background plane (which depends upon the lens strength).
Nevertheless, the probability that
a source is a strong lens should be an increasing function of
strong-lensing cross section and hence of predicted
Einstein radius.  Figure~\ref{ratefig} shows this
effect for the SLACS targets with well-measured SDSS velocity dispersions.
We see a rise from a $\approx$20\% confirmation at
a predicted $b$ of $0\farcs 5$ up to a $\approx$100\%
confirmation rate at a predicted $b$ of $2\arcsec$.

In some cases, spatially resolved
integral-field unit (IFU) spectroscopy of SLACS targets is available
from a separate survey program using the Magellan and Gemini telescopes.
The details of this IFU survey, along with narrow-band images
extracted from the IFU data cubes showing the spatial morphology
of the background line emission, are presented by \citet{bb_07}.  
A subset of these IFU data were also presented in Paper~I, showing
how the spatial coincidence between putative lensed features in the
\textsl{HST} imaging and high-redshift emission-line flux in the
IFU data can solidify the strong-lens hypothesis.
The full list of SLACS targets with Magellan and Gemini
IFU spectroscopy is presented in Table~\ref{ifutab}, along with
brief comments on the implications of the IFU data for the interpretation
of the \textsl{HST} imaging.  A separate program to obtain VLT
IFU spectroscopy (at lower spatial resolution but higher SNR)
is described in \citet{czoske_2321}.

\begin{table*}
\footnotesize
\begin{center}
\caption{\label{ifutab} Summary of Magellan/Gemini
integral-field spectroscopic
evidence for/against lensing in SLACS systems}
\begin{tabular}{ll}
\hline \hline
System Name & Comments on IFU$+$\textsl{HST} Data \\
\hline
SDSSJ0037$-$0942 & Clear coincidence of IFU line emission and \textsl{HST} lensed features \\
SDSSJ0044$+$0113 & Clear coincidence of IFU line emission and \textsl{HST} lensed features \\
SDSSJ0737$+$3216 & Clear coincidence of IFU line emission and \textsl{HST} lensed features \\
SDSSJ0956$+$5100 & Low-SNR IFU line emission coincident with \textsl{HST} lensed features \\
SDSSJ1029$+$6115 & Lensed galaxy rotation curve in IFU data; \textsl{HST} imaging ambiguous. \\
SDSSJ1155$+$6237 & IFU shows emission-line source not multiply imaged, despite multiple \textsl{HST}. \\
SDSSJ1259$+$6134 & Low-SNR possible lensing features in IFU and \textsl{HST}; very inconclusive \\
SDSSJ1402$+$6321 & Clear coincidence of line emission and \textsl{HST} lensed features \\
SDSSJ1416$+$5136 & Clear coincidence of IFU line emission and \textsl{HST} lensed features \\
SDSSJ1630$+$4520 & Clear coincidence of IFU line emission and \textsl{HST} lensed features \\
SDSSJ1702$+$3320 & Low-SNR possible lensing features in IFU and \textsl{HST}; inconclusive \\
SDSSJ2238$-$0754 & Clear coincidence of line emission and \textsl{HST} lensed features \\
SDSSJ2302$-$0840 & Clear lensed ring in IFU data; \textsl{HST} imaging ambiguous. \\
SDSSJ2321$-$0939 & Clear coincidence of line emission and \textsl{HST} lensed features \\
\hline
\end{tabular}
\end{center}
\end{table*}

\subsection{Mass modeling}
\label{lensmod}

Here we describe our strong-lens mass modeling procedure and results.
Construction of a successful strong gravitational lens model
is necessary both to solidify the lensing hypothesis in a candidate
lens system and to make the lens-mass measurements of scientific interest.
Lens models must simultaneously describe the distribution of
light in the un-lensed background ``source plane'' and the distribution of
mass in the foreground ``lens plane'' that generates the gravitational potential
through which the source plane is viewed.

For all systems classified as grade-A lenses,
we fit the putative lensed images with
a singular isothermal ellipsoid (SIE) lens model
\citep{kormann_sie, kassiola_kovner, keeton_kochanek}.  The SIE
model consists of similar concentric and aligned elliptical
isodensity contours with axis ratio $q_{\mathrm{SIE}}$.
In the circular ($q=1$) limit, the projected surface density of the SIE
falls off as $\Sigma \propto R^{-1}$ in two dimensions.
The model is parameterized by its
angular Einstein radius $b$, which is related to
the physical mass model through
\begin{equation}
b = 4 \pi {{\sigma_{\mathrm{SIE}}^2} \over {c^2}}
{D_{LS} \over D_S}~.
\label{b_sie}
\end{equation}
Here, $\sigma_{\mathrm{SIE}}$ is a velocity-dispersion parameter
and $D_{LS}$ and $D_S$ are cosmological angular-diameter distances from lens to
source and observer to source respectively.
As in previous SLACS papers, we adopt the intermediate-axis normalization
of \citet{kormann_sie}, whereby the mass within
a given isodensity contour remains constant at fixed $b$ for changing
axis ratio $q_{\mathrm{SIE}}$.  We model the lensed background galaxies
as either single or multiple Gaussian or S\'{e}rsic 
ellipsoid components
as necessary to obtain a good fit.  The center of the mass model is constrained
to be coincident with the center of the lens-galaxy light profile.
Initial trial values for the lens-model Einstein radius and axis ratio
are taken from the separation of the candidate lensed images and from
the ellipticity of the light profile.
The model lensed image is generated by ray-tracing through the analytic
SIE mass model to view the parameterized source galaxy model, and
subsequently convolved with the ACS PSF.
All model parameters (lens and source) are adjusted manually to
approximately match the data, and are then optimized using \texttt{MPFIT}.
The final outcome is a set of lens-model and source-component parameters,
along with a model for the lensed image configuration.
This parametric source-plane technique (also employed by B07 and \citealt{marshall_0737})
can be contrasted with the pixellated source-plane techniques for
modeling resolved optical sources described by \citet{warren_dye_03},
\citet{tk04}, \citet{wayth_lensview}, and Paper~III\@.  We employ the
parameterized strategy for its simplicity and ease of implementation,
and for the robustness of the resulting aperture-mass measurements.
Future work will apply the pixellated source-plane method
to the full SLACS lens sample.

Figure~\ref{prettypics} and Table~\ref{lenstable} in
Appendix~\ref{dataapp} present
the best-fit lens-model images and parameters that result
from this modeling procedure.
With the exception of the systems listed in Table~\ref{specialnotes}
of Appendix~\ref{dataapp} (which all involve complicating factors as described),
the SIE analysis yields successful models of
the lensed surface-brightness distribution.  In certain cases we
see data--model mismatch at the level of detailed features, as is
to be expected given the parameterization of the source-plane
surface-brightness distribution in terms of Gaussian and S\'ersic
ellipsoids.  This outcome confirms the essential
validity of our visual classifications: the trained eye is
in fact quite good at ``mental modeling'' and hypothesis testing
in strong lensing.

It is worth noting that the SIE lens modeling
succeeds with the peak of the mass distribution
constrained to be coincident with the peak of the luminosity
profile.  This is consistent with stellar mass being the dominant
contributor to the gravitational field in the central
kiloparsecs, and requires that the dark-matter component and
any significant gas mass be well aligned with the stellar spheroid.
Furthermore, this coincidence requires that the SLACS
lenses must be located at or very near to the center of mass
of any environmental overdensities (groups/clusters) in which
they may be located.
We can quantify the extent of the average mass-light centroid coincidence
by continuing the lens-model optimization while freeing the mass
centroid to move in position.
For this analysis, we identify a subset of 32 grade-A lenses that
are either complete or nearly complete ``Einstein rings'' with relatively
high SNR lensed features,
which we refer to as the ``ring subset''
and which are identified in Table~\ref{lenstable}
(in Appendix~\ref{dataapp}).
Since the lensed images in this subset extend through a
large range in azimuth about the lens center,
the mass centroids of the lens models are especially well
constrained.
We find an RMS shift of the mass centroid of
$0\farcs044$---approximately one native ACS pixel.
Such shifts are probably small enough to be consistent with no shift
at all, given the many accumulated sources of minor uncertainty.
Converting the shifts
to physical scales at the lens redshifts, the RMS mass centroid
shift is 140 parsecs.  As a fraction of the measured Einstein radii,
the RMS shift is 3.5\%.  The quantitative implications of this
positional mass--light alignment will be explored in a future
SLACS publication.

In most scientific applications of strong lensing, measured
Einstein radii are of primary interest, providing direct determinations
of the enclosed mass.  In the case of Einstein ring images or symmetric
quadruple-image lenses, this aperture-mass measurement is nearly
independent of the radial density profile of the adopted lens model \citep{kochanek_91}.
When the lensed image configuration is significantly
asymmetric, the Einstein radius parameter measured from
the data becomes somewhat dependent on the assumed mass model
\citep[e.g.,][]{rkk_03}.
To assess the magnitude of this effect in the SLACS sample,
we also fit all SIE-modeled systems with light-traces-mass (LTM) lens models
derived from the B-spline galaxy models.  We use the deconvolved
B-spline ellipsoid model with no multipole dependence, since the higher-order
models needed to produce the best residual images are in some cases
unstable under deconvolution.  We compute the lensing deflection of
the LTM models directly from the deconvolved B-spline model images using
fast Fourier techniques \citep[e.g.,][]{wayth_lensview},
and take the overall mass-to-light ratio for each system
as a free parameter analogous to the Einstein radius parameter of the SIE model.
We also include an external shear and its position angle
as free parameters, in order to allow for angular
degrees of freedom analogous to
the free axis ratio and position-angle parameters of the SIE,
which are necessary in order to obtain reasonable fits \citep[e.g.][]{kks_97}.
In a comparative sense, this can give a slight advantage
to the LTM over the SIE models, since the former
can model both an internal quadrupole
moment (through the fixed ellipticity of the light profile) and an external
quadrupole moment (through the shear).  In the majority of cases, however, the
best-fit SIE and LTM model images for the lensed features are
visually indistinguishable from one another.  We note however
that the results of Paper~III (based on combined lensing and dynamical
models), Paper~IV (based on combined strong- and weak-lensing analysis),
Paper~VI (for the double Einstein ring SDSSJ0946$+$1006),
and Paper~VII (based on homologous ensemble strong-lensing analysis) strongly
favor the SIE radial mass-density profile over the LTM profile for the
SLACS lens sample
(also see \citealt{kt02, kt03, tk02, tk03, tk04},
\citealt{rkk_03}; \citealt{rusin_kochanek_05}).
We convert the fitted LTM mass-to-light ratios into LTM Einstein radii
by determining the radial
position at which the lensing deflection of the best-fit LTM mass model
exactly matches the radial offset from the lens center
in the circular limit.  The LTM mass model parameters are given in
Table~\ref{lenstable}, and the model images can be seen along with the SIE
model images in Figure~\ref{prettypics}
(all within Appendix~\ref{dataapp}).  These LTM lens models are
used alongside the SIE models in Paper~VII to assess the dependence
of the derived physical scaling relations upon the assumed form of the
lensing mass model.

\section{Measurement comparisons and errors}
\label{mcomp}

In this section we compare our mass and light
parameter measurements to values obtained for the same systems
through different procedures.  This provides both a sanity
check and a more realistic sense of the measurement
errors associated with the individual parameters.
Formal statistical errors can be obtained
from the parameter covariance matrices (evaluated at the
best-fit parameter values in the case of non-linear fits),
but these estimates only account for the contribution
of photon-noise and read-noise to the error budget, and they
only apply to the idealized case where the true luminosity
and mass distributions under study are exactly of the forms
described by the parameterized models.
Though the system-by-system uncertainties in all quantities of interest
will in general depend upon the details of the lensed
image configuration in that system and upon the depth of observation,
subsequent work will benefit from the determination of
the typical realistic uncertainty across the sample
in each measured parameter. In particular, the empirical
scaling-relation analyses of Paper~VII employ the estimates
of characteristic errors that we derive here.
Table~\ref{errortab} presents a summary of the formal
statistical errors and the adopted empirical errors
derived from the analysis of this section.

\begin{table*}[t]
\begin{center}
\caption{\label{errortab} Formal and empirical measurement-error estimates}
\begin{tabular}{lrr}
\hline \hline
Measured Quantity & Formal Statistical Error & Adopted Empirical Error \\
\hline
Einstein radius $b$ & 0.2\%  & 2\% \\
Mass axis ratio $q_{\mathrm{SIE}}$ & 0.005 & 0.05 \\
Mass position angle & $<1^{\circ}$ & $6^{\circ}$ ($2^{\circ}$ for ring subset) \\
De Vauc.\ magnitude & 0.001--0.002\,mag & 0.03\,mag \\
Effective radius $R_e$ & $<$0.2\% & 3.5\% \\
Velocity dispersion & 7\% & 7\% \\
\hline
\end{tabular}
\end{center}
\end{table*}

\subsection{Mass model parameters}

\begin{figure}
\plotone{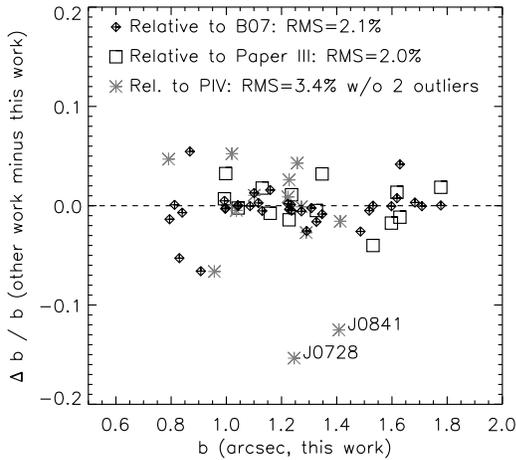}
\caption{\label{re_comp}
Fractional difference between SIE Einstein radii $b$ from the analyses
of other SLACS papers and this work.}
\end{figure}

First we compare our SIE Einstein radius measurements with those
measured for subsets of SLACS lenses in Paper~III (14 systems in common)
and Paper~IV (13 systems in common), as well
as with the measurements made for B07 (34 systems in common).
The lens modeling of Paper~III and Paper~IV was carried out with a regularized pixellated
source plane as opposed to the multi-S\'{e}rsic models of this work.
The models of B07, meanwhile, were parameterized in the same manner as in
the current work, but were fitted directly to the native pixel data of single
Snapshot exposures, rather than to the rectified (and in some cases combined)
frames used in this work.  We also note that the values published in
Table~1 of Paper~IV
reflect an error in the conversion from major-axis to
intermediate-axis conventions, and should be divided by the square-root
of the mass axis ratio to provide for a proper comparison to the
values of this paper.  The corrections are small, and we have
verified that the results and conclusions of Paper~IV are not significantly
altered by the change.  Figure~\ref{re_comp} shows the fractional
difference between SIE Einstein radii measured by different methods for
the same systems, as a function of SIE Einstein radius $b$.
The RMS fractional differences are 2\% for Paper~III and B07 relative
to the current work, with no significant systematic bias.
Relative to this work, the values of Paper~IV exhibit a larger
6\% fractional scatter, though this reduces to 3\% with no
significant systematic offset if the
two outlying systems J0728 and J0841 are excluded.
With regard to these two systems: J0728 shows complex lensing
morphology that may admit qualitatively different lens-model
interpretations, while J0841 is a highly asymmetric double-image
lens, for which the measured Einstein radius can be more
significantly degenerate with a combination of mass
axis ratio and position angle.
In subsequent analyses, we will adopt a 2\% RMS fractional error
as our best estimate of the uncertainty on measured Einstein
radii.  For comparison, the median formal statistical error for
the 63 Einstein-radius measurements given in this paper is 0.2\%.

We also wish to determine the actual error in the lensing measurements
of the projected mass axis ratio and major-axis position angle
through comparison of current measurements to the SIE models of B07.
For the minor-to-major projected mass axis ratio $q_{\mathrm{SIE}}$
(which ranges between 0 and 1), we find an RMS difference of 0.05
with no significant systematic bias.  We will adopt 0.05
as our typical parameter uncertainty, which
contrasts with the much smaller median formal statistical error in
$q_{\mathrm{SIE}}$ of 0.005.
Comparing position-angle measurements, we find a mean difference (B07 minus
this work) of $0.8^{\circ} \pm 1.0^{\circ}$---consistent with
no systematic misalignment---with an RMS difference of 5.7$^{\circ}$,
after rejecting three outlier systems with large position-angle differences
between the two works.  Even in these outlier cases
(J0935$-$0003, J1204$+$0358, and J1403$+$0006), the Einstein-radius
measurements between the two works agree to within 5\%, a fact which
highlights the relative robustness of the Einstein radius among lens parameter
measurements.
For the ring subset defined in \S\ref{lensmod}---for which the
angular mass properties are especially well constrained---the
RMS position-angle difference
(B07 minus this work) is a much smaller $2.0^{\circ}$.
Thus we see that a $2^{\circ}$--$6^{\circ}$ RMS statistical uncertainty
applies for the measured mass position angles, though we recognize
that catastrophic outliers may creep in.  For comparison, the median formal statistical
error in the measured mass position angles is less than 1$^{\circ}$.

\subsection{Surface-brightness model parameters}

Next we compare multiple F814W de Vaucouleurs surface photometry measurements
for the same target galaxies.   Perhaps the
best check of purely ``statistical'' (though not photon-counting)
photometric errors is obtained through comparison of magnitudes measured for
the 16 systems with both Snapshot (discovery) and
full-orbit (follow-up) observations through the same filter.
In this case, the mean offset (snapshot minus full-orbit) is
$-0.01$ magnitude, with an RMS offset of $0.03$ magnitude.
For comparison, the formal statistical error estimates for
these measurements are at the level of one to two
milli-magnitudes.  We will adopt this value of 0.03 magnitudes as our
photometric error estimate for all systems.  Though the full-orbit
measurements should arguably be given smaller errors, we do
not wish to over-weight the 43 lens systems with full-orbit F814W
photometry relative to the 20 with only Snapshot measurements
(and in any event, the dominant magnitude errors are not
set by the observation depth).

The measurements
made and published previously in the SLACS series have
used slightly different model-fitting procedures,
and the resulting dispersion in values provides a further
check on our levels of statistical and systematic confidence.
The closest comparison is to the Snapshot photometry of
15 systems in Paper~II,
for which the reduction, masking, and fitting procedures were
most similar (though not identical) to the current methods.
We find a mean offset (Paper~II minus this work) of $-0.026$ magnitudes
and an RMS difference of $0.047$ magnitudes.  Comparing next to 21
photometric values published in Paper~IV,
we find a mean offset (Paper~IV minus this work)
of $-0.013$ and an RMS difference of $0.2$
magnitudes.\footnote{The apparent F814W magnitude
for SDSSJ1023$+$4230 as published in Paper~IV should read 16.93.
The Paper~IV \textit{absolute} magnitude of this galaxy
is correct as printed.}
The Paper~IV values were taken from models fitted to a significantly
smaller region ($24\arcsec$ to a side, versus the $54\arcsec$
to a side used in this work), and the Paper~IV masking procedure
was fully automated whereas this work applies manual masks in the
inner $14\arcsec \times 14\arcsec$.  Paper~IV measurements also included
a free diskiness/boxiness parameter whereas the models of the current
work are pure ellipsoids.  Thus we
interpret the scatter between these two sets of
magnitudes as evidence of the well-known
effect that de Vaucouleurs magnitude of a galaxy depends both on the
galaxy itself and upon the fitting procedure used, due to
departures of the real galaxy from the simple de Vaucouleurs ellipsoid form.
While this could perhaps be mitigated by the use of the S\'{e}rsic
model, the extrapolated flux in the low-surface brightness
wings of the S\'{e}rsic model
is highly dependent upon the S\'{e}rsic index $n$, and becomes quite a
large fraction of the total model flux when $n$ becomes large.
Use of S\'{e}rsic magnitudes would also greatly complicate comparison
with other studies based upon de Vaucouleurs photometry.  Thus we
work with de Vaucouleurs magnitudes here and in the scaling-relation
analyses of Paper~VII\@.

We also compare de Vaucouleurs effective (half-light) radius
measurements---taken from the same model fits as the magnitudes---from
multiple measurement procedures.  Comparing Snapshot to full-orbit
measurements as above, we find a mean fractional offset (Snapshot minus
full orbit) of 1\% and an RMS difference of 3.5\%; we will adopt
this value as our empirical error estimate going forward.
The median formal fractional statistical error in the effective radius measurements,
by comparison, is less than two tenths of one percent.
Comparing the Snapshot
measurements of this work to those of Paper~II (converting the
latter from a major-axis to an intermediate-axis convention),
we find a mean offset (Paper~II minus this work) of 0.1\% and an RMS difference of 5\%.
Comparing to Paper~IV values, we find a mean difference (Paper~IV minus this work)
of 0.5\% and an RMS difference of 25\%.  The significant scatter between current
and Paper~IV values we again attribute to the significant differences in
analysis procedures.  Finally, we compare the values measured in the current work
to $i$-band de Vaucouleurs effective radii from the SDSS photometric
database (converting SDSS values from major axis to intermediate axis).
Excluding the 6 systems with multiple foreground-galaxy multiplicities,
and rejecting a further 6 outlier systems (leaving a sample
of 119 total), we find a mean offset
(SDSS minus this work) of $-0.7$\% and an RMS difference of 12\%.
Again, we note that the de Vaucouleurs effective radius depends
largely upon the analysis details.
Similar scatter in the precise determination of
effective radii has been found by \citep{kelson_2000_i} and
\citet{treu_2001_ii}.

Since errors on the de Vaucouleurs magnitudes and effective radii
are significantly correlated, we also derive an empirical error
in the effective surface brightness, proportional to the model luminosity
divided by the square of the model effective radius.
From the comparison of Snapshot to full-orbit measurements of
this quantity, we find an RMS fractional difference of 4.5\%,
which we will adopt as our empirical uncertainty in the measured effective
surface brightnesses.

\subsection{Velocity-dispersion measurements}

The stellar velocity dispersion measurements that we present
in this paper and use extensively in Paper~VII are measured
from SDSS spectroscopic data by the Princeton/MIT analysis
pipeline.\footnote{\texttt{http://spectro.princeton.edu/}}
The SDSS spectrograph fibers sample
a seeing-convolved circular spatial aperture of 3$\arcsec$
in diameter centered on the target galaxies.
The median seeing is $\approx\!\! 1\farcs4$; the physical scale of
the fiber diameter is about 10\,kpc at a redshift of $z=0.2$.
Velocity dispersions are measured by fitting a linear combination
of stellar templates to the observed galaxy data in pixel space, weighted
using the estimated observational errors and masking pixels at
common emission-line wavelengths.  All templates are shifted
together by a free velocity-shift parameter (initialized using the primary
galaxy-redshift value), and broadened by a single Gaussian kernel
described by a free velocity-dispersion parameter (in addition
to broadening by the fixed spectrograph resolution).
A grid of trial velocity-dispersion and velocity-shift parameters
is explored, and the corresponding $\chi^2$ values are
mapped out by optimizing the stellar template coefficients
linearly at each grid point.
The best-fit velocity dispersion is derived at the $\chi^2$ minimum of
a quadratic fit to those points near the minimum in the grid values.
The stellar templates themselves are derived from a
principal-components analysis of the original ELODIE library of high-resolution
stellar spectra \citep{elodie}, keeping the 24 most
significant eigenvectors from 886 of the 908 stars
The ELODIE spectra cover the rest-frame wavelength
range 4100--6800\AA\@.  For analyses that make use of
these velocity dispersions, we only consider the subset of
foreground galaxies with median spectral SNR of 10 or greater
per 69\,km\,s$^{-1}$ pixel over this wavelength range.
This prevents the inclusion of velocity-dispersion data points
with excessively (or catastrophically) large errors.

Unlike most other measurements reported in
this work, the SDSS velocity dispersions
are generally in an SNR regime where the dominant
contribution to the uncertainty is due to the statistics of
photon counting.  We test for any further random
uncertainty by comparing the velocity dispersions measured from
the same data by two different revisions of the Princeton/MIT
pipeline---one run following SDSS-DR4 and one
following SDSS-DR6.  For all simple early-type
systems observed by SLACS with sufficient SDSS spectroscopic SNR to permit
a measurement, the velocity-dispersion values from the two
different runs are consistent, with a reduced $\chi^2$ of 0.77
across the sample.  We thus adopt the formal statistical error estimates
directly, though we limit the fractional error estimate for any one
system to a minimum of 5\% in view of the systematic errors associated with
possible mismatch in the stellar templates used in the measurement.
Where necessary, we adopt 7\% as a single overall value for
the uncertainty in all the velocity dispersion measurements, though this value
will necessarily be an underestimate of some errors and an overestimate
of others.

\section{Control-sample tests}
\label{compsample}

As discussed in Papers I~and~II, our ability to generalize deductions
from the SLACS lens sample to the larger population of early-type
galaxies requires an understanding of our selection procedure
and of any possible biases that procedure may introduce.
In a nutshell, the SLACS target selection is for the following:
\begin{enumerate}
\item A quiescent spectrum of the target SDSS galaxy,
\item The presence of higher redshift emission lines in the SDSS spectrum, and
\item Appreciable lensing cross section as estimated from redshifts
   and stellar velocity dispersions.
\end{enumerate}
For the resulting \textit{lens} sample, an additional condition is
\begin{enumerate}
\setcounter{enumi}{3}
\item The detection of strongly lensed features in \textsl{HST} imaging.
\end{enumerate}
Our approach here will be similar to that employed in Papers I~and~II:
we replicate conditions 1 and 3 by constructing comparison samples
for each target from the SDSS database by identifying galaxies with (roughly) the
same redshift, spectral quiescence, and velocity dispersion.
The massive data volume of
the SDSS spectroscopic database allows us to construct our comparison
samples by directly matching observed quantities, thus limiting
sensitivity to additional corrections.
Furthermore---and unlike the analyses of Papers~I and~II---we also
require the comparison-sample galaxies to have nearly equal
effective radii to the corresponding SLACS targets.  The combination
of velocity-dispersion and effective-radius constraints ensures that
the comparison samples should be located at the same point on the
fundamental plane as the SLACS galaxies.
If conditions 2 and 4 work to make the SLACS target sample significantly
biased or un-representative, this should manifest as
a biased distribution in magnitudes for the SLACS galaxies
relative to their control samples.

Our recipe for constructing the comparison samples is summarized as follows.
We work with the SDSS DR6 photometric and spectroscopic catalog \citep{dr6} so
as to have the largest possible parent sample, with all data reduced
by a single version of the SDSS photometric and spectroscopic pipelines.
We select the overall parent sample by requiring a Princeton/MIT
SDSS spectroscopic pipeline classification of \texttt{GALAXY}
and a rest-frame H$\alpha$ equivalent-width measurement
of either less than 4\AA\
in value or less than 2 sigma in significance---conditions
likewise required for SLACS target selection, with the exception
of several late-type lenses and lens candidates.  We also require that
the best-fit spectroscopic pipeline template describe the
spectrum with a reduced $\chi^2$ of no more than 3, since a
poor spectral model fit prevents the significant detection of
higher-redshift emission lines in the model-subtracted residual spectrum.
We impose a minimum median spectral SNR of
10 per pixel over the rest-frame range 4100--6800\AA\@.
This SNR value is computed
from the observed-frame \texttt{SN\_MEDIAN} reported by the
Princeton/MIT pipeline through
an empirical redshift-dependent conversion determined
from the SDSS spectroscopy of the SLACS targets:
the approximate rest-frame median SNR per pixel
is given by \texttt{SN\_MEDIAN}
$ +~20.4 \times z_{\mathrm{lens}} - 1.24$.  Finally, we
require that the SDSS $r$-band de Vaucouleurs effective radius
be well measured, and that the magnitudes in all five
SDSS filters also be well measured.  For each SLACS target,
we identify the subset of this parent sample
within $dz = \pm 0.01$ of the SLACS target redshift.  We then identify
a further subset with effective radii within $\pm 7.5$\% of
the SLACS target value and with velocity dispersions in
an interval containing the SLACS target value.
The width of the velocity bin is set to 15\% of the measured
SLACS target velocity-dispersion value, and
the bin center is chosen so as to give an equal number of
comparison galaxies at higher and lower dispersion than the target.
As noted in Paper~II, a balancing of this sort is necessary due to
the steepness of the velocity-dispersion function.  For the
confirmed SLACS lenses, the resulting comparison samples have
from 26 to 2996 galaxies, with a median sample size of 666.
These figures exclude the high velocity-dispersion lens SDSSJ0935$-$0003,
which has only two comparison-sample galaxies (which are both brighter
than the lens, by 0.1 and 0.5 magnitudes respectively).

\begin{figure}
\plotone{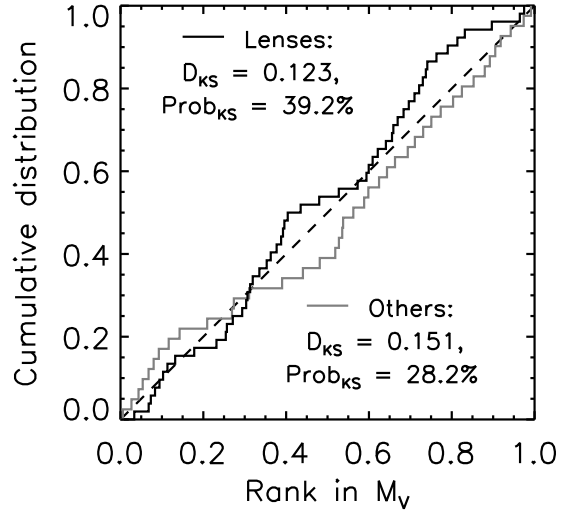}
\caption{\label{kstest}
Kolmogorov-Smirnov (K-S) tests of the rank of 52 A-grade
SLACS lenses and 41 other SLACS targets
within the distributions of absolute magnitude $M_V$ of their individual
SDSS comparison samples.  Only systems with single multiplicity,
early-type morphology,
and well-measured SDSS velocity dispersions are included.
The null-hypothesis distribution---corresponding to a representative
drawing of the SLACS systems from spectroscopically
comparable galaxies in the SDSS---is given by
the linear cumulative distribution shown with a dashed line.
The K-S $D$ statistic values are given, along with
the probability of random occurrence of an equal or greater $D$
value under the null hypothesis.}
\end{figure}

With the comparison samples in hand for each SLACS target,
we examine the distribution of SLACS magnitudes
within these samples.
We use SDSS values for the control samples as well as for the SLACS
targets, so as to avoid complications of photometric
zero-point matching.
We reduce the SDSS fluxes of the modeled lenses
by a percentage corresponding to the contribution of the lensed
images to the total $I$-band flux within a seeing-convolved circle
of radius 3$\arcsec$, as measured from the \textsl{HST}-ACS data.
For reference, we find an offset between ACS F814W
de~Vaucouleurs magnitude and SDSS $i$-band magnitude
for the SLACS lenses
given by $i_{\mathrm{SDSS}} - I_{814} = 0.17$,
with an RMS scatter of 0.19 magnitude.
We compute absolute $V$-band magnitudes from the SDSS fluxes
using distance moduli for our assumed cosmology and
$k$ corrections computed using the \texttt{SDSS2BESSELL}
procedure of the \texttt{kcorrect} software \citep{kcorrect}.
We also apply our adopted luminosity-evolution correction,
though it makes a difference of only $\pm 0.01$ magnitude
over the redshift width of the comparison-sample bins.
For each SLACS target with a well-measured velocity dispersion,
we then determine its rank within the cumulative distributions of
absolute magnitude for its comparison sample.  The rank values
range from $0.5/N_{\mathrm{samp}}$ to $1 - (0.5/N_{\mathrm{samp}})$,
where $N_{\mathrm{samp}}$ is the number of galaxies in the
comparison sample including the SLACS target.  If the targets
are drawn in a representative fashion from their parent
samples, these ranks should be distributed uniformly
between 0 and 1---a proposition we can test with the
Kolmogorov-Smirnov (K-S) formalism.  Figure~\ref{kstest} shows
this K-S test of the absolute-magnitude rank distribution
for the 52 early-type SLACS A-grade lenses with well-measured
SDSS velocity dispersions (``lenses'', excluding
SDSSJ0935$-$0003), as well as for the 41 other SLACS early-type
target systems with single multiplicity and similarly well-measured
velocity dispersions (``others'').  The ``lenses'' are consistent
with their parent samples at the 39.2\% level, while
the ``others'' are
consistent with their parent samples at the 28.2\% level.
A two-sample K-S tests show that the two SLACS target populations
(``lenses'' and ``others'') are consistent with one another
in their distributions at the 39.0\% level.
We can also test for any systematic bias as a function of
intrinsic lens-galaxy properties by testing for correlations
between the absolute-magnitude rank of lenses within
their control samples and their position within the $R_e$-$\sigma_{e2}$
plane.  If any such correlations were present, then the
SLACS lenses would define a biased FP relative to their
control samples.  In fact there are no such significant correlations:
the linear correlation coefficient between magnitude-rank
and effective radius (in physical units) is
$r = -0.080$, and the correlation with velocity-dispersion
is $r = -0.088$.  The correlation of magnitude rank with the
product $\sigma_{e2}^2 R_e$ (proportional to the ``dynamical
mass'' of the lens) is $r = -0.065$.
From these tests we conclude that the SLACS lenses and
other targets are statistically consistent with having been drawn
at random from the parent SDSS galaxy population with similar
spectroscopic properties.

\section{Summary and conclusions}
\label{sumcon}

We have presented an up-to-date catalog of
the largest single confirmed strong gravitational
lens sample to date, from the ACS data set
of the SLACS Survey.
The catalog includes 63 ``grade-A'' strong galaxy-galaxy lens
systems complete
with lens and source redshifts, F814W lens-galaxy photometry,
gravitational lens models, and (in most cases)
stellar velocity dispersions.
Such a large and high-quality
lens sample serves as a further proof-of-concept for
the spectroscopic discovery channel, and
provides a unique resource for the quantitative study
of massive early-type galaxies.
Many of the most immediate implications of our measurements
for the structure and physical scaling relations of early-type
galaxies are explored further in Paper~VII.

We have described the details of the image-analysis
and parameterized lens-modeling techniques that we use to make
mass and luminosity measurements from the \textsl{HST}-ACS
imaging data.  Our analysis demonstrates that simple singular isothermal
ellipsoid and light-traces-mass (plus external shear) lens models,
combined with multiple Gaussian or S\'{e}rsic ellipsoid models
of the lensed background galaxies, can reproduce the lensed
image configurations in great detail.
(More detailed modeling with pixellated source-plane
surface-brightness distributions is currently
being conducted to further
reduce the level of systematic residuals and to extract
all strong-lensing information.)
The current models imply a precise
positional alignment of the peaks of the mass and light
distributions in the foreground lensing galaxies.  We have also presented
a realistic empirical analysis of the characteristic
errors associated with the various measurements reported
in this work, which are in general much larger than
purely random-statistical considerations would indicate.
Finally, we have demonstrated that the SLACS lens sample
is statistically consistent with having been drawn
at random from a parent population of similar
galaxies from the SDSS, a conclusion that supports the generalization
of SLACS results to the massive early-type galaxy population in general.

The strong lensing measurements presented in this work afford
a unique opportunity to test the results of numerical
simulations of galaxy formation, merging, and evolution.
This is due to the fact that strong lensing measures total
mass directly, in a nearly model-independent sense,
and without need for modeling of stellar populations
and luminosity evolution.  One can envision a particularly simple
test as follows.  For a particular formation and merger-progenitor
scenario, one can select simulated galaxies corresponding to each
of the observed lens galaxies by identifying those with identical (or nearly
identical) effective radii of the stellar tracer component (regardless
of luminosity) and identical projected aperture masses within
the physical Einstein radius of the lens.  The line-of-sight
velocity dispersions would then be computed for the simulated
counterparts, and compared to the observed velocity dispersions
of the lens galaxies.  Through the level of agreement between these
predicted and observed velocity dispersions across the full range
of relevant scales, various formation
scenarios could in principle be distinguished from one another.
This amounts to a test of whether or not the simulated galaxies
define the same mass plane---as defined
in B07 and discussed further in Paper~VII, in analogy to the
fundamental plane---but through
direct comparison with the data,
rather than through the comparison of scaling-relation coefficients.

The main limitations to further quantitative study of the
SLACS lens sample are due to
(1) the observational error in the velocity dispersions derived from
SDSS spectroscopy, and (2) the lack of high-resolution multi-color
imaging of the full sample.  To address the first limitation,
follow-up spectroscopy of SLACS lenses is being
pursued at the Keck and VLT observatories \citep{czoske_2321}.
This spectroscopy additionally affords spatial resolution, allowing a direct
measurement of the stellar kinematics within fixed physical
apertures.  The second limitation is being addressed though continued
\textsl{HST} imaging of confirmed lenses in multiple bands,
which will allow quantitative study of the stellar
populations within the SLACS lens galaxies and their lensed
background source galaxies \citep{marshall_0737}.

\acknowledgments

ASB, TT, LVEK, RG, and LAM acknowledge the support
and hospitality of the Kavli
Institute for Theoretical Physics
at UCSB, where a significant part of
this work was completed.
ASB thanks G. Dobler for valuable discussion
related to this work.
TT acknowledges support
from the NSF through CAREER award NSF-0642621 and from the Sloan Foundation through a
Sloan Research Fellowship.  He is also supported by a Packard fellowship.
LVEK is supported
in part through an NWO-VIDI program subsidy (project number 639.042.505).
He also acknowledges the continuing support by the European Community's
Sixth Framework Marie Curie Research Training Network Programme,
Contract No. MRTN-CT-2004-505183 (``ANGLES'')\@.
The work of LAM was carried out at Jet Propulsion Laboratory, California
Institute of Technology under a contract with NASA\@.

Support for \textsl{HST} programs \#10174, \#10494,
\#10587, \#10798, and \#10886 was provided by NASA through a grant from
the Space Telescope Science Institute,
which is operated by the Association of Universities for
Research in Astronomy, Inc., under NASA contract NAS 5-26555.
Please see \textsl{HST} data acknowledgment on title page.

This work has made extensive use of the Sloan Digital Sky Survey database.  Funding for the SDSS and SDSS-II has been provided by the Alfred P. Sloan Foundation, the Participating Institutions, the National Science Foundation, the U.S. Department of Energy, the National Aeronautics and Space Administration, the Japanese Monbukagakusho, the Max Planck Society, and the Higher Education Funding Council for England. The SDSS Web Site is \texttt{http://www.sdss.org/}.

The SDSS is managed by the Astrophysical Research Consortium for the Participating Institutions. The Participating Institutions are the American Museum of Natural History, Astrophysical Institute Potsdam, University of Basel, University of Cambridge, Case Western Reserve University, University of Chicago, Drexel University, Fermilab, the Institute for Advanced Study, the Japan Participation Group, Johns Hopkins University, the Joint Institute for Nuclear Astrophysics, the Kavli Institute for Particle Astrophysics and Cosmology, the Korean Scientist Group, the Chinese Academy of Sciences (LAMOST), Los Alamos National Laboratory, the Max-Planck-Institute for Astronomy (MPIA), the Max-Planck-Institute for Astrophysics (MPA), New Mexico State University, Ohio State University, University of Pittsburgh, University of Portsmouth, Princeton University, the United States Naval Observatory, and the University of Washington.

\appendix

\section{Data tables and image figures}
\label{dataapp}

\begin{figure*}[h]
\plotone{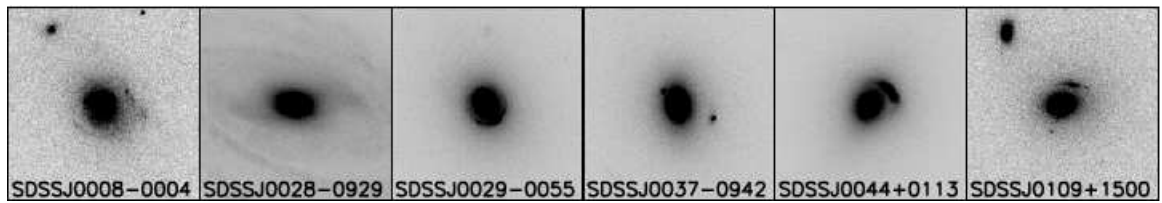}
\caption{\label{allimage}
\textsl{HST}-ACS WFC imaging through the F814W filter of
SLACS targets.  Images are $10 \arcsec \times 10 \arcsec$, with North
up and East to the left.  Cosmic-ray pixels in single-exposure images have
been replaced with smoothed image values.  Grayscale is linear from
$-0.25 X$ (white) to $X$ (black), where $X$ is the 98th percentile flux level
in the image.  A version of this figure with all 131 target panels is
available through
the electronic edition of the \textsl{Astrophysical Journal}, or through
the website of the first author.
}
\end{figure*}

\setlength{\tabcolsep}{3pt}

\begin{table}[h]
\begin{center}
\caption{\label{obstable} SLACS HST-ACS target observational data}
\begin{tabular}{ccccccccccccc}
\hline \hline
RA/Dec & Plate-MJD- & ~ & ~ & $I_{\mathrm{814}}$ & $I_{814}$ & $L_{V555}$ & $R_e$ & $L_{e2} /$ & $B/A$ & $PA$ & $\sigma_{\mathrm{SDSS}}$ & Classifi- \\
(J2000) & FiberID & $z_{\mathrm{FG}}$ & $z_{\mathrm{BG}}$ & (obs.) & extin. & ($10^9 L_{\odot}$) & ($\arcsec$) & $L_{\mathrm{deV}}$ & (deV) & ($^{\circ}$) & (km\,s$^{-1}$) & cation \\
\hline
000802.96$-$000408.2 & 0669-52559-156 & 0.4400 & 1.1924 & 18.65d & 0.12 & \phn 86.7 &  1.71 & 0.313 & 0.83 &  27.3 & \nodata & E-S-A\phn \\
002817.87$-$092934.3 & 0653-52145-590 & 0.0565 & 0.7146 & 13.75s & 0.07 & \phn 92.8 & 15.21 & 0.346 & 0.48 &  78.9 & 147$\pm$7\phn & L-S-X\phn \\
002907.77$-$005550.5 & 0391-51782-088 & 0.2270 & 0.9313 & 17.09d & 0.04 & \phn 76.3 &  2.16 & 0.310 & 0.84 &  26.6 & 229$\pm$18 & E-S-A\phn \\
003753.21$-$094220.1 & 0655-52162-392 & 0.1955 & 0.6322 & 16.26s & 0.06 & 120.5 &  2.19 & 0.326 & 0.73 &  11.4 & 279$\pm$14 & E-S-A\phn \\
004402.90$+$011312.6 & 0393-51794-456 & 0.1196 & 0.1965 & 15.73s & 0.04 & \phn 68.8 &  2.61 & 0.321 & 0.76 & 151.3 & 266$\pm$13 & E-S-A\phn \\
010933.73$+$150032.5 & 0422-51811-508 & 0.2939 & 0.5248 & 17.75s & 0.11 & \phn 77.2 &  1.38 & 0.320 & 0.78 & 104.0 & 251$\pm$19 & E-S-A\phn \\
015758.94$-$005626.1 & 0700-52199-020 & 0.5132 & 0.9243 & 18.64d & 0.05 & 117.3 &  1.06 & 0.310 & 0.69 &  69.2 & \nodata & E-S-A\phn \\
021652.54$-$081345.3 & 0668-52162-428 & 0.3317 & 0.5235 & 16.93d & 0.07 & 206.4 &  2.67 & 0.312 & 0.79 &  81.2 & 333$\pm$23 & E-S-A\phn \\
025245.21$+$003958.4 & 0807-52295-614 & 0.2803 & 0.9818 & 18.04d & 0.15 & \phn 55.8 &  1.39 & 0.317 & 0.94 &  97.2 & 164$\pm$12 & E-S-A\phn \\
033012.14$-$002051.9 & 0810-52672-252 & 0.3507 & 1.0709 & 18.16d & 0.16 & \phn 82.3 &  1.20 & 0.306 & 0.77 & 109.6 & 212$\pm$21 & E-S-A\phn \\
035458.47$-$064842.8 & 0464-51908-310 & 0.1301 & 0.3808 & 15.90s & 0.14 & \phn 76.9 &  3.76 & 0.316 & 0.88 &   9.1 & 160$\pm$8\phn & E-S-X\phn \\
040535.41$-$045552.4 & 0465-51910-406 & 0.0753 & 0.8098 & 16.45s & 0.21 & \phn 15.8 &  1.36 & 0.320 & 0.69 &  20.3 & 160$\pm$8\phn & E-S-A\phn \\
072804.95$+$383525.7 & 1733-53047-154 & 0.2058 & 0.6877 & 16.74d & 0.12 & \phn 91.2 &  1.78 & 0.316 & 0.74 &  67.0 & 214$\pm$11 & E-S-A\phn \\
073728.45$+$321618.6 & 0541-51959-145 & 0.3223 & 0.5812 & 17.04d & 0.08 & 177.8 &  2.82 & 0.312 & 0.85 & 104.1 & 338$\pm$17 & E-S-A\phn \\
074251.84$+$345001.9 & 0542-51993-386 & 0.0853 & 0.7390 & 17.02s & 0.11 & \phn 11.0 &  1.77 & 0.314 & 0.97 & 124.5 & 165$\pm$10 & E-S-X\phn \\
075834.68$+$303443.3 & 1061-52641-256 & 0.1156 & 0.5013 & 16.05s & 0.10 & \phn 50.3 &  1.37 & 0.320 & 0.81 & 108.0 & 191$\pm$10 & E-S-B\phn \\
080240.82$+$450452.7 & 0436-51883-633 & 0.1423 & 0.4523 & 16.16s & 0.09 & \phn 70.3 &  2.71 & 0.315 & 0.80 &  86.0 & 244$\pm$12 & E-S-X\phn \\
080358.21$+$453655.6 & 0439-51877-333 & 0.1313 & 0.2938 & 16.90s & 0.15 & \phn 31.6 &  1.07 & 0.315 & 0.36 &  69.5 & 228$\pm$12 & L-S-X\phn \\
080858.78$+$470638.9 & 0438-51884-555 & 0.2195 & 1.0251 & \nodata & 0.09 & \nodata & \nodata & \nodata & \nodata & \nodata & \nodata & E-M-A* \\
081931.93$+$453444.8 & 0441-51868-108 & 0.1943 & 0.4462 & 17.07s & 0.08 & \phn 57.6 &  1.98 & 0.319 & 0.78 &  40.4 & 225$\pm$15 & E-S-B\phn \\
082242.32$+$265243.5 & 1267-52932-253 & 0.2414 & 0.5941 & 16.99d & 0.05 & \phn 95.4 &  1.82 & 0.312 & 0.74 &  87.0 & 259$\pm$15 & E-S-A\phn \\
084128.81$+$382413.7 & 0828-52317-012 & 0.1159 & 0.6567 & 15.34d & 0.06 & \phn 94.6 &  4.21 & 0.318 & 0.58 &  92.9 & 225$\pm$11 & L-S-A\phn \\
084706.89$+$031822.6 & 0564-52224-542 & 0.1192 & 0.4146 & 16.80s & 0.05 & \phn 25.7 &  1.91 & 0.320 & 0.69 & 120.1 & 199$\pm$12 & E-S-X\phn \\
090315.19$+$411609.1 & 1200-52668-398 & 0.4304 & 1.0645 & 17.95d & 0.03 & 144.7 &  1.78 & 0.308 & 0.89 &   1.6 & \nodata & E-S-A\phn \\
090319.52$+$313951.2 & 1590-52974-622 & 0.2711 & 0.5494 & 16.77d & 0.04 & 150.0 &  3.03 & 0.319 & 0.67 & 147.8 & 258$\pm$15 & E-S-B\phn \\
091053.11$+$052023.2 & 1193-52652-232 & 0.2706 & 1.0741 & \nodata & 0.08 & \nodata & \nodata & \nodata & \nodata & \nodata & \nodata & E-M-B\phn \\
091205.31$+$002901.2 & 0472-51955-429 & 0.1642 & 0.3239 & 15.57d & 0.05 & 156.4 &  3.87 & 0.330 & 0.67 &  11.7 & 326$\pm$16 & E-S-A\phn \\
091244.31$+$413637.0 & 1200-52668-588 & 0.0646 & 0.1377 & 15.85s & 0.04 & \phn 17.3 &  1.24 & 0.337 & 0.65 &  22.7 & 218$\pm$11 & E-S-X\phn \\
092559.35$+$081411.8 & 1302-52763-012 & 0.1345 & 0.2251 & 17.66s & 0.09 & \phn 15.6 &  1.44 & 0.322 & 0.83 &  67.2 & \nodata & E-S-X\phn \\
093425.13$+$603423.5 & 0486-51910-241 & 0.1011 & 0.2440 & 16.37s & 0.06 & \phn 27.5 &  1.65 & 0.313 & 0.56 &  48.4 & 208$\pm$10 & E-S-X\phn \\
093543.93$-$000334.8 & 0476-52314-177 & 0.3475 & 0.4670 & 16.75s & 0.06 & 268.5 &  4.24 & 0.311 & 0.90 & 145.2 & 396$\pm$35 & E-S-A\phn \\
093600.77$+$091335.8 & 1303-53050-078 & 0.1897 & 0.5880 & 16.52d & 0.07 & \phn 90.5 &  2.11 & 0.308 & 0.81 & 145.3 & 243$\pm$12 & E-S-A\phn \\
094656.68$+$100652.8 & 1305-52757-503 & 0.2219 & 0.6085 & 17.09d & 0.05 & \phn 73.2 &  2.35 & 0.316 & 0.96 &  10.3 & 263$\pm$21 & E-S-A\phn \\
095320.42$+$520543.7 & 0902-52409-577 & 0.1315 & 0.4673 & 17.26s & 0.01 & \phn 20.1 &  1.22 & 0.319 & 0.88 &  45.5 & 229$\pm$19 & E-S-X\phn \\
095519.72$+$010144.4 & 0268-51633-336 & 0.1109 & 0.3159 & 16.97s & 0.05 & \phn 18.9 &  1.09 & 0.304 & 0.39 & 108.6 & 192$\pm$13 & L-S-A\phn \\
095629.78$+$510006.6 & 0902-52409-068 & 0.2405 & 0.4699 & 16.68d & 0.02 & 122.7 &  2.19 & 0.311 & 0.73 & 147.7 & 334$\pm$17 & E-S-A\phn \\
095900.96$+$441639.4 & 0942-52703-499 & 0.2369 & 0.5315 & 16.90d & 0.02 & \phn 97.6 &  1.98 & 0.317 & 0.87 &  55.9 & 244$\pm$19 & E-S-A\phn \\
095944.07$+$041017.0 & 0572-52289-495 & 0.1260 & 0.5350 & 16.92d & 0.05 & \phn 25.9 &  1.39 & 0.298 & 0.60 &  58.4 & 197$\pm$13 & E-S-A\phn \\
101622.86$+$385903.3 & 1427-52996-461 & 0.1679 & 0.4394 & 16.71d & 0.03 & \phn 56.7 &  1.46 & 0.320 & 0.85 &  63.3 & 247$\pm$13 & E-S-A\phn \\
102026.54$+$112241.1 & 1598-53033-353 & 0.2822 & 0.5530 & 17.21d & 0.06 & 110.5 &  1.59 & 0.319 & 0.79 & 106.6 & 282$\pm$18 & E-S-A\phn \\
102332.26$+$423001.8 & 1359-53002-418 & 0.1912 & 0.6960 & 16.77d & 0.03 & \phn 70.1 &  1.77 & 0.314 & 0.85 & 167.5 & 242$\pm$15 & E-S-A\phn \\
102551.32$-$003517.5 & 0272-51941-151 & 0.1589 & 0.2764 & 15.41s & 0.12 & 181.2 &  4.94 & 0.312 & 0.76 & 112.3 & 264$\pm$13 & E-S-X\phn \\
102922.94$+$042001.8 & 0576-52325-433 & 0.1045 & 0.6154 & 16.13d & 0.06 & \phn 36.7 &  1.56 & 0.315 & 0.52 & 127.9 & 210$\pm$11 & E-S-A\phn \\
102927.53$+$611505.3 & 0772-52375-140 & 0.1574 & 0.2512 & 16.06s & 0.02 & \phn 88.8 &  2.73 & 0.360 & 0.83 &   3.1 & 228$\pm$14 & E-S-B\phn \\
103235.84$+$532234.9 & 0905-52643-100 & 0.1334 & 0.3290 & 17.05d & 0.03 & \phn 25.5 &  0.81 & 0.306 & 0.44 & 136.5 & 296$\pm$15 & L-S-A\phn \\
103904.22$+$051335.8 & 0577-52367-571 & 0.0668 & 0.3627 & 15.38s & 0.05 & \phn 28.7 &  2.40 & 0.323 & 0.87 &  59.0 & 190$\pm$10 & E-S-X\phn \\
103957.78$+$093351.0 & 1240-52734-507 & 0.2212 & 0.5612 & \nodata & 0.05 & \nodata & \nodata & \nodata & \nodata & \nodata & \nodata & E-M-B\phn \\
\hline
\end{tabular}
\tablecomments{Plate-MJD-Fiber constitute a unique SDSS spectrum identifier.
Redshifts $z_{\mathrm{FG}}$ and $z_{\mathrm{BG}}$ are for
foreground and background galaxies respectively, as measured from SDSS data:
$z_{\mathrm{FG}}$ values are taken directly from the SDSS database,
while $z_{\mathrm{BG}}$ values are measured as described in \citet{bolton_speclens}.
Apparent magnitudes $I_{814}$ are from \textsl{HST}-ACS de Vaucouleurs models, and are
quoted in the AB system \textit{without} correction for Galactic extinction.
Magnitudes are measured from either 420-s Snapshot
exposures (``s'' for ``snap'') or full-orbit multi-exposure images
(``d'' for ``deep'').  $I$-band Galactic dust extinction
values based on \citet{sfd_dust} maps are given separately,
and should be subtracted
from observed magnitudes to give dust-corrected magnitudes.
Rest-frame luminosities $L_{V555}$ are as computed from $I_{814}$
with corrections for Galactic extinction, evolution, $k$-correction,
and cosmological distance modulus as described in the text,
assuming an absolute solar AB magnitude of $V_{555,\odot} = 4.83$.
Effective radii $R_e$ are measured from de Vaucouleurs image models,
and quoted at the intermediate axis.  $L_{e2}/L_{\mathrm{deV}}$ gives
ratio of luminosity within $R_e / 2$ as determined from B-spline
models to total de Vaucouleurs model luminosity.  $B/A$ gives ratio
of minor to major axes for the de Vaucouleurs image models.
$PA$ gives de Vauc.\ major-axis
position angles measured E from N.  Velocity dispersions
$\sigma_{\mathrm{SDSS}}$ are uncorrected for aperture effects.
Reported errors are limited to a minimum of $0.05 \sigma_{\mathrm{SDSS}}$.
No $\sigma_{\mathrm{SDSS}}$
values are reported for systems whose median SNR is less than
10 over the range of rest-frame wavelengths used for the fit,
or for systems with multiple foreground galaxies.
``Classification'' column gives codes denoting (1) foreground-galaxy
morphology, (2) foreground-galaxy
multiplicity, and (3) status of system as a lens based on available data.
Morphology is coded by ``E'' for early-type (elliptical and S0),
``L'' for late-type (Sa and later), and ``U'' for unclassified
(galaxies that cannot be unambiguously classed as early- or late-type
based on the \textsl{HST}-ACS data).  Multiplicity is coded by
``S'' for single and ``M'' for multiple.  Lens status is coded by
``A'' for systems with clear and convincing evidence of multiple
imaging, ``B'' for systems with strong evidence of multiple imaging
but insufficient SNR for definite conclusion and/or modeling,
and ``X'' for all other systems (non-lenses and non-detections).
Systems marked as ``A*'' are definite lenses, but are not modeled
for reasons specified in Table~\ref{specialnotes}.
}
\end{center}
\end{table}

\addtocounter{table}{-1}
\begin{table}[h]
\begin{center}
\caption{\label{dummylabel} (continued)}
\begin{tabular}{ccccccccccccc}
\hline \hline
RA/Dec & Plate-MJD- & ~ & ~ & $I_{\mathrm{814}}$ & $I_{814}$ & $L_{V555}$ & $R_e$ & $L_{e2} /$ & $B/A$ & $PA$ & $\sigma_{\mathrm{SDSS}}$ & Classifi- \\
(J2000) & FiberID & $z_{\mathrm{FG}}$ & $z_{\mathrm{BG}}$ & (obs.) & extin. & ($10^9 L_{\odot}$) & ($\arcsec$) & $L_{\mathrm{deV}}$ & (deV) & ($^{\circ}$) & (km\,s$^{-1}$) & cation \\
\hline
104606.93$+$415116.1 & 1361-53047-077 & 0.1025 & 0.7584 & 16.61d & 0.02 & \phn 22.0 &  1.03 & 0.308 & 0.36 &  49.8 & 191$\pm$10 & L-S-X\phn \\
110024.39$+$532913.9 & 1011-52652-175 & 0.3171 & 0.8581 & 17.18s & 0.02 & 143.6 &  2.24 & 0.314 & 0.58 & 103.0 & \nodata & E-S-A\phn \\
110308.21$+$532228.2 & 1011-52652-156 & 0.1582 & 0.7353 & 16.43d & 0.02 & \phn 63.7 &  1.95 & 0.330 & 0.46 &  45.5 & 196$\pm$12 & U-S-A\phn \\
110646.15$+$522837.8 & 1011-52652-007 & 0.0955 & 0.4069 & 15.52s & 0.02 & \phn 51.4 &  1.68 & 0.324 & 0.63 &  57.3 & 262$\pm$13 & E-S-A\phn \\
110817.70$+$025241.3 & 0509-52374-471 & 0.1368 & 0.3105 & 17.19s & 0.08 & \phn 24.7 &  1.17 & 0.322 & 0.86 & 146.1 & 178$\pm$14 & E-S-B\phn \\
111250.60$+$082610.4 & 1221-52751-028 & 0.2730 & 0.6295 & 17.22s & 0.06 & 101.9 &  1.50 & 0.328 & 0.77 & 137.5 & 320$\pm$20 & E-S-A\phn \\
111739.60$+$053414.0 & 0835-52326-571 & 0.2285 & 0.8230 & 17.11s & 0.12 & \phn 81.4 &  2.20 & 0.308 & 0.72 &  43.9 & 277$\pm$19 & E-S-B\phn \\
113405.89$+$602713.5 & 0952-52409-524 & 0.1528 & 0.4742 & 16.44s & 0.02 & \phn 59.1 &  2.02 & 0.325 & 0.83 & 155.0 & 239$\pm$12 & E-S-A\phn \\
113636.14$+$042625.0 & 0837-52642-039 & 0.1282 & 0.5341 & 16.97s & 0.04 & \phn 25.4 &  0.88 & 0.330 & 0.81 & 123.9 & 258$\pm$14 & E-S-X\phn \\
114052.69$+$564044.5 & 1312-52781-311 & 0.0674 & 0.2968 & 15.78s & 0.02 & \phn 19.6 &  1.92 & 0.321 & 0.69 & 145.9 & 163$\pm$9\phn & L-S-X\phn \\
114257.35$+$100111.8 & 1226-52734-306 & 0.2218 & 0.5039 & 17.10d & 0.10 & \phn 75.8 &  1.91 & 0.314 & 0.89 &  95.4 & 221$\pm$22 & E-S-A\phn \\
114329.64$-$014430.0 & 0328-52282-535 & 0.1060 & 0.4019 & 14.96d & 0.03 & 108.5 &  4.80 & 0.337 & 0.80 & 118.7 & 269$\pm$13 & E-S-A\phn \\
115208.97$+$005431.0 & 0284-51943-452 & 0.1062 & 0.1590 & 16.90s & 0.04 & \phn 18.5 &  0.86 & 0.328 & 0.58 & 123.5 & 235$\pm$14 & E-S-X\phn \\
115310.79$+$461205.3 & 1446-53080-211 & 0.1797 & 0.8751 & 17.20d & 0.04 & \phn 41.9 &  1.16 & 0.323 & 0.90 &   2.9 & 226$\pm$15 & E-S-A\phn \\
115510.06$+$623722.4 & 0777-52320-501 & 0.3751 & 0.6690 & 17.61s & 0.03 & 141.3 &  2.88 & 0.323 & 0.77 & 176.9 & \nodata & E-S-X\phn \\
115905.46$+$544738.3 & 1018-52672-279 & 0.0818 & 0.2695 & 15.74d & 0.02 & \phn 30.6 &  1.90 & 0.318 & 0.69 & 107.3 & 231$\pm$12 & E-S-X\phn \\
120324.89$+$023301.1 & 0517-52024-352 & 0.1644 & 0.4380 & 16.59s & 0.05 & \phn 61.2 &  2.70 & 0.312 & 0.50 &  67.2 & 209$\pm$11 & L-S-X\phn \\
120444.07$+$035806.4 & 0842-52376-208 & 0.1644 & 0.6307 & 16.84s & 0.04 & \phn 48.1 &  1.47 & 0.316 & 0.97 & 132.1 & 267$\pm$17 & E-S-A\phn \\
120540.44$+$491029.4 & 0969-52442-134 & 0.2150 & 0.4808 & 16.56d & 0.04 & 110.4 &  2.59 & 0.314 & 0.72 & 158.3 & 281$\pm$14 & E-S-A\phn \\
121158.75$+$455036.6 & 1370-53090-427 & 0.1110 & 0.3170 & 15.63d & 0.02 & \phn 63.6 &  2.89 & 0.322 & 0.75 & 107.6 & 231$\pm$12 & E-S-X\phn \\
121340.58$+$670829.0 & 0493-51957-145 & 0.1229 & 0.6402 & 15.60d & 0.03 & \phn 81.1 &  3.23 & 0.326 & 0.77 &  20.0 & 292$\pm$15 & E-S-A\phn \\
121826.70$+$083050.3 & 1625-53140-415 & 0.1350 & 0.7172 & 15.74d & 0.03 & \phn 87.2 &  3.18 & 0.321 & 0.72 &  50.5 & 219$\pm$11 & E-S-A\phn \\
124426.03$+$011146.8 & 0291-51928-528 & 0.0725 & 0.5600 & 15.21s & 0.03 & \phn 39.2 &  2.83 & 0.320 & 0.70 & 106.9 & 172$\pm$9\phn & L-S-X\phn \\
125028.26$+$052349.1 & 0847-52426-549 & 0.2318 & 0.7953 & 16.70d & 0.05 & 115.4 &  1.81 & 0.310 & 0.97 & 114.8 & 252$\pm$14 & E-S-A\phn \\
125050.52$-$013531.7 & 0337-51997-460 & 0.0871 & 0.3526 & 15.14s & 0.04 & \phn 61.7 &  2.93 & 0.317 & 0.72 & 125.3 & 246$\pm$12 & U-S-A* \\
125135.71$-$020805.2 & 0337-51997-480 & 0.2243 & 0.7843 & 17.25s & 0.04 & \phn 63.8 &  2.61 & 0.299 & 0.51 &  39.5 & \nodata & L-S-A\phn \\
125919.05$+$613408.6 & 0783-52325-279 & 0.2334 & 0.4488 & 16.85s & 0.02 & \phn 98.9 &  1.81 & 0.314 & 0.79 &  96.1 & 253$\pm$16 & E-S-A* \\
131326.70$+$050657.2 & 0851-52376-344 & 0.1438 & 0.3385 & 17.10s & 0.06 & \phn 29.4 &  0.86 & 0.311 & 0.45 &  74.9 & 221$\pm$17 & L-S-B\phn \\
133045.53$-$014841.6 & 0910-52377-503 & 0.0808 & 0.7115 & 16.99s & 0.07 & \phn 9.8 &  0.89 & 0.315 & 0.46 & 103.6 & 185$\pm$9\phn & E-S-B\phn \\
134308.25$+$602755.0 & 0786-52319-236 & 0.1198 & 0.3199 & 16.30s & 0.03 & \phn 40.6 &  2.07 & 0.310 & 0.52 & 153.1 & 178$\pm$10 & L-S-X\phn \\
134309.22$+$605209.7 & 0786-52319-193 & 0.0343 & 0.0880 & 13.64s & 0.03 & \phn 35.8 &  4.91 & 0.322 & 0.49 &   5.1 & 206$\pm$10 & E-S-X\phn \\
140228.21$+$632133.5 & 0605-52353-503 & 0.2046 & 0.4814 & 16.33d & 0.03 & 122.1 &  2.70 & 0.316 & 0.77 &  70.8 & 267$\pm$17 & E-S-A\phn \\
140329.49$+$000641.4 & 0302-51688-354 & 0.1888 & 0.4730 & 17.11s & 0.08 & \phn 52.8 &  1.46 & 0.317 & 0.81 & 110.5 & 213$\pm$17 & E-S-A\phn \\
141622.34$+$513630.4 & 1045-52725-464 & 0.2987 & 0.8111 & 17.57d & 0.02 & \phn 87.5 &  1.43 & 0.326 & 0.76 &  23.4 & 240$\pm$25 & E-S-A\phn \\
142015.85$+$601914.8 & 0788-52338-605 & 0.0629 & 0.5351 & 15.08d & 0.03 & \phn 32.8 &  2.06 & 0.326 & 0.57 & 111.5 & 205$\pm$10 & E-S-A\phn \\
143004.10$+$410557.1 & 1349-52797-406 & 0.2850 & 0.5753 & 16.87d & 0.02 & 149.4 &  2.55 & 0.309 & 0.79 & 120.7 & 322$\pm$32 & E-S-A\phn \\
143039.86$+$511530.9 & 1046-52460-448 & 0.1337 & 0.4503 & 16.33s & 0.02 & \phn 48.9 &  1.81 & 0.333 & 0.68 &  74.1 & 206$\pm$10 & L-S-X\phn \\
143213.34$+$631703.8 & 0499-51988-005 & 0.1230 & 0.6643 & 15.16d & 0.03 & 122.5 &  5.85 & 0.307 & 0.96 & 107.2 & 199$\pm$10 & L-S-A\phn \\
143609.50$+$493927.3 & 1046-52460-025 & 0.1225 & 0.3145 & 16.32s & 0.04 & \phn 42.4 &  2.13 & 0.312 & 0.71 &  12.9 & 212$\pm$12 & E-S-X\phn \\
143627.54$-$000029.2 & 0306-51637-035 & 0.2852 & 0.8049 & 17.24s & 0.07 & 112.2 &  2.24 & 0.315 & 0.75 & 151.3 & 224$\pm$17 & E-S-A\phn \\
144319.62$+$030408.2 & 0587-52026-205 & 0.1338 & 0.4187 & 17.06s & 0.06 & \phn 26.1 &  0.94 & 0.320 & 0.62 &  61.1 & 209$\pm$11 & E-S-A\phn \\
144858.24$-$011614.6 & 0920-52411-607 & 0.1474 & 0.7807 & 16.65s & 0.10 & \phn 48.2 &  1.39 & 0.302 & 0.41 &  34.4 & 187$\pm$10 & L-S-X\phn \\
145128.19$-$023936.4 & 0921-52380-293 & 0.1254 & 0.5203 & 16.09d & 0.16 & \phn 61.0 &  2.48 & 0.315 & 0.98 &  40.6 & 223$\pm$14 & E-S-A\phn \\
145218.94$-$005820.2 & 0309-51994-298 & 0.1770 & 0.5131 & 17.28s & 0.08 & \phn 39.3 &  0.85 & 0.321 & 0.77 & 120.9 & 193$\pm$11 & E-S-X\phn \\
151505.14$+$612848.3 & 0611-52055-626 & 0.2421 & 0.3800 & 17.31s & 0.03 & \phn 70.3 &  1.20 & 0.327 & 0.66 & 173.0 & 212$\pm$25 & E-S-X\phn \\
152009.08$-$003457.3 & 0313-51673-306 & 0.1140 & 0.3954 & 16.88s & 0.11 & \phn 23.0 &  1.61 & 0.324 & 0.59 &  30.7 & 196$\pm$16 & L-S-X\phn \\
152444.37$-$005209.1 & 0924-52409-527 & 0.1524 & 0.7323 & 17.39s & 0.28 & \phn 31.0 &  1.62 & 0.313 & 0.82 &  54.8 & 150$\pm$22 & E-S-X\phn \\
152506.70$+$332747.4 & 1387-53118-532 & 0.3583 & 0.7173 & 17.11d & 0.04 & 204.0 &  2.90 & 0.316 & 0.61 & 135.4 & 264$\pm$26 & E-S-A\phn \\
152524.63$+$011401.7 & 0313-51673-523 & 0.1294 & 0.6269 & 16.68s & 0.09 & \phn 35.3 &  1.59 & 0.319 & 0.85 &  58.1 & 158$\pm$10 & E-S-X\phn \\
153150.07$-$010545.7 & 0314-51641-124 & 0.1596 & 0.7439 & 16.08s & 0.26 & 112.6 &  2.50 & 0.320 & 0.68 & 143.5 & 279$\pm$14 & E-S-A\phn \\
153530.38$-$003852.3 & 0315-51663-259 & 0.1613 & 0.6585 & 16.81s & 0.21 & \phn 55.9 &  1.23 & 0.325 & 0.60 &  98.2 & 254$\pm$15 & E-S-X\phn \\
153711.26$+$412554.6 & 1679-53149-628 & 0.1423 & 0.6811 & 17.02d & 0.04 & \phn 30.3 &  2.07 & 0.316 & 0.84 &   1.7 & 204$\pm$14 & E-S-X\phn \\
153812.92$+$581709.8 & 0615-52347-594 & 0.1428 & 0.5312 & 16.66s & 0.03 & \phn 42.0 &  1.58 & 0.311 & 0.82 & 153.5 & 189$\pm$12 & E-S-A\phn \\
154100.77$+$413058.7 & 1053-52468-275 & 0.1423 & 0.5033 & 16.84d & 0.05 & \phn 36.2 &  1.15 & 0.320 & 0.42 &  64.0 & 215$\pm$11 & L-S-X\phn \\
154731.22$+$572000.0 & 0617-52072-561 & 0.1883 & 0.3958 & 16.25s & 0.02 & 109.4 &  2.53 & 0.317 & 0.89 & 156.8 & 254$\pm$13 & E-S-X\phn \\
155003.12$+$525846.7 & 0618-52049-458 & 0.0491 & 0.7396 & 15.23s & 0.03 & \phn 17.1 &  2.04 & 0.335 & 0.75 & 108.9 & 202$\pm$10 & E-S-B\phn \\
160453.49$+$335546.2 & 1418-53142-599 & 0.0786 & 0.3500 & 15.36d & 0.05 & \phn 40.7 &  2.59 & 0.319 & 0.63 &  97.2 & 228$\pm$11 & E-S-B\phn \\
161437.74$+$452253.3 & 0814-52443-510 & 0.1779 & 0.8113 & 16.83s & 0.02 & \phn 56.9 &  2.58 & 0.316 & 0.90 &  60.5 & 182$\pm$13 & E-S-B\phn \\
161843.10$+$435327.4 & 0815-52374-337 & 0.1989 & 0.6657 & \nodata & 0.03 & \nodata & \nodata & \nodata & \nodata & \nodata & \nodata & E-M-A* \\
162132.99$+$393144.6 & 1172-52759-318 & 0.2449 & 0.6021 & 16.81s & 0.01 & 113.2 &  2.14 & 0.312 & 0.73 & 142.9 & 236$\pm$20 & E-S-A\phn \\
162746.45$-$005357.6 & 0364-52000-084 & 0.2076 & 0.5241 & 16.91d & 0.18 & \phn 85.1 &  1.98 & 0.312 & 0.85 &   6.9 & 290$\pm$15 & E-S-A\phn \\
163028.16$+$452036.3 & 0626-52057-518 & 0.2479 & 0.7933 & 16.79d & 0.01 & 118.4 &  1.96 & 0.318 & 0.84 &  71.7 & 276$\pm$16 & E-S-A\phn \\
\hline
\end{tabular}
\end{center}
\end{table}

\addtocounter{table}{-1}
\begin{table}[h]
\begin{center}
\caption{\label{dummylabel2} (continued)}
\begin{tabular}{ccccccccccccc}
\hline \hline
RA/Dec & Plate-MJD- & ~ & ~ & $I_{\mathrm{814}}$ & $I_{814}$ & $L_{V555}$ & $R_e$ & $L_{e2} /$ & $B/A$ & $PA$ & $\sigma_{\mathrm{SDSS}}$ & Classifi- \\
(J2000) & FiberID & $z_{\mathrm{FG}}$ & $z_{\mathrm{BG}}$ & (obs.) & extin. & ($10^9 L_{\odot}$) & ($\arcsec$) & $L_{\mathrm{deV}}$ & (deV) & ($^{\circ}$) & (km\,s$^{-1}$) & cation \\
\hline
163339.26$-$001256.2 & 0348-51671-234 & 0.0702 & 0.2060 & 15.81s & 0.17 & \phn 23.9 &  2.28 & 0.323 & 0.52 & 169.7 & 215$\pm$11 & U-S-X\phn \\
163602.62$+$470729.6 & 0627-52144-464 & 0.2282 & 0.6745 & 17.03s & 0.04 & \phn 81.5 &  1.68 & 0.321 & 0.78 & 102.2 & 231$\pm$15 & E-S-A\phn \\
170013.98$+$622109.7 & 0349-51699-043 & 0.1228 & 0.3584 & 16.52s & 0.05 & \phn 35.3 &  1.53 & 0.314 & 0.72 & 118.5 & 192$\pm$10 & E-S-X\phn \\
170216.76$+$332044.8 & 0973-52426-464 & 0.1785 & 0.4357 & 16.10s & 0.04 & 113.2 &  3.66 & 0.313 & 0.78 & 116.3 & 256$\pm$14 & E-S-B\phn \\
170603.69$+$330400.9 & 0974-52427-127 & 0.1682 & 0.7736 & 16.85d & 0.04 & \phn 50.5 &  1.38 & 0.321 & 0.79 &  26.3 & 225$\pm$12 & E-S-B\phn \\
171723.13$+$573948.2 & 0355-51788-542 & 0.1144 & 0.5748 & 16.02s & 0.06 & \phn 48.9 &  2.08 & 0.315 & 0.77 & 145.6 & 227$\pm$11 & E-S-X\phn \\
171837.40$+$642452.2 & 0352-51789-563 & 0.0899 & 0.7366 & \nodata & 0.06 & \nodata & \nodata & \nodata & \nodata & \nodata & \nodata & E-M-A* \\
211112.27$-$003826.5 & 0986-52443-256 & 0.1933 & 0.4761 & \nodata & 0.15 & \nodata & \nodata & \nodata & \nodata & \nodata & \nodata & E-M-B\phn \\
211949.65$-$074201.7 & 0639-52146-142 & 0.1704 & 0.5262 & 16.76s & 0.40 & \phn 78.0 &  1.91 & 0.311 & 0.64 & 139.4 & 207$\pm$17 & E-S-B\phn \\
212151.12$+$120312.9 & 0730-52466-327 & 0.1434 & 0.4862 & 16.92s & 0.12 & \phn 36.3 &  1.61 & 0.313 & 0.59 &  66.9 & 194$\pm$12 & L-S-B\phn \\
214154.68$-$000112.3 & 0989-52468-035 & 0.1380 & 0.7127 & 16.83s & 0.10 & \phn 35.8 &  1.81 & 0.300 & 0.37 &  88.4 & 181$\pm$14 & L-S-A* \\
220218.32$-$084648.0 & 0717-52468-165 & 0.1613 & 0.5011 & 17.13s & 0.07 & \phn 36.6 &  1.13 & 0.312 & 0.34 &  12.7 & 231$\pm$12 & L-S-X\phn \\
220956.93$-$075447.9 & 0718-52206-475 & 0.1112 & 0.2148 & 17.58s & 0.09 & \phn 11.2 &  0.71 & 0.317 & 0.49 & 178.3 & 229$\pm$16 & E-S-X\phn \\
222537.34$+$125957.6 & 0737-52518-119 & 0.3103 & 0.6571 & 18.09d & 0.15 & \phn 66.2 &  0.74 & 0.325 & 0.75 & 136.7 & 248$\pm$24 & E-S-X\phn \\
223840.20$-$075456.0 & 0722-52224-442 & 0.1371 & 0.7126 & 16.20d & 0.07 & \phn 61.2 &  2.33 & 0.315 & 0.74 & 138.3 & 198$\pm$11 & E-S-A\phn \\
224155.71$+$122814.0 & 0739-52520-054 & 0.0998 & 0.7173 & 15.92s & 0.07 & \phn 40.9 &  4.55 & 0.350 & 0.54 & 164.8 & 176$\pm$13 & L-S-X\phn \\
230053.15$+$002238.0 & 0677-52606-520 & 0.2285 & 0.4635 & 17.07d & 0.10 & \phn 83.1 &  1.83 & 0.321 & 0.80 &  85.7 & 279$\pm$17 & E-S-A\phn \\
230220.18$-$084049.5 & 0725-52258-463 & 0.0901 & 0.2224 & 15.53s & 0.07 & \phn 47.4 &  2.25 & 0.325 & 0.80 & 169.1 & 237$\pm$12 & E-S-A* \\
230321.72$+$142217.9 & 0743-52262-304 & 0.1553 & 0.5170 & 16.10d & 0.35 & 112.9 &  3.28 & 0.321 & 0.64 &  36.7 & 255$\pm$16 & E-S-A\phn \\
232120.93$-$093910.3 & 0645-52203-517 & 0.0819 & 0.5324 & 14.66s & 0.05 & \phn 84.6 &  4.11 & 0.313 & 0.78 & 127.9 & 249$\pm$12 & E-S-A\phn \\
234111.57$+$000018.7 & 0682-52525-594 & 0.1860 & 0.8070 & 16.36d & 0.05 & \phn 98.7 &  3.15 & 0.318 & 0.59 &  78.8 & 207$\pm$13 & E-S-A\phn \\
234728.08$-$000521.3 & 0684-52523-311 & 0.4169 & 0.7145 & 17.89s & 0.06 & 145.3 &  1.40 & 0.309 & 0.71 &  16.5 & \nodata & E-S-B\phn \\
\hline
\end{tabular}
\end{center}
\end{table}

\begin{table}[h]
\footnotesize
\begin{center}
\caption{\label{lenstable} SLACS HST-ACS grade-A strong lens model parameters}
\begin{tabular}{cccccccccccc}
\hline \hline
System Name & $b_{\mathrm{SIE}}$ & $q$ & $PA$ & $L_{\mathrm{Ein,SIE}}$ & $b_{\mathrm{LTM}}$ & $\gamma_{\mathrm{ext}}$ & $PA_{\gamma}$ & $L_{\mathrm{Ein,LTM}}$ & ~ & Ring & Good \\
(SDSS\ldots) & ($\arcsec$) & (SIE) & (SIE) & $ / L_{\mathrm{deV}}$ & ($\arcsec$) & (LTM) & (LTM) & $/ L_{\mathrm{deV}}$ & $N_{\mathrm{src}}$ & Subset? & $\sigma_{\mathrm{SDSS}}$? \\
\hline
J0008$-$0004 & 1.16 & 0.70 &  35.2 & 0.393 & 1.14 & 0.09 &  37.6 & 0.387 & 3 & No & No \\
J0029$-$0055 & 0.96 & 0.89 &  25.4 & 0.284 & 0.95 & 0.01 &  33.1 & 0.282 & 2 & Yes & Yes \\
J0037$-$0942 & 1.53 & 0.84 &  15.9 & 0.404 & 1.52 & 0.01 &  67.1 & 0.401 & 2 & No & Yes \\
J0044$+$0113 & 0.79 & 0.66 &   7.4 & 0.218 & 0.76 & 0.12 &  19.4 & 0.211 & 2 & No & Yes \\
J0109$+$1500 & 0.69 & 0.55 &  99.8 & 0.321 & 0.68 & 0.07 &  83.8 & 0.317 & 1 & No & Yes \\
J0157$-$0056 & 0.79 & 0.72 & 102.6 & 0.401 & 0.67 & 0.24 & 103.1 & 0.362 & 3 & No & No \\
J0216$-$0813 & 1.16 & 0.79 &  73.3 & 0.283 & 1.15 & 0.03 &  78.6 & 0.282 & 3 & No & Yes \\
J0252$+$0039 & 1.04 & 0.93 & 106.2 & 0.441 & 1.03 & 0.01 &  99.2 & 0.439 & 3 & Yes & Yes \\
J0330$-$0020 & 1.10 & 0.81 & 113.2 & 0.459 & 1.04 & 0.07 & 113.9 & 0.443 & 3 & No & Yes \\
J0405$-$0455 & 0.80 & 0.72 &  21.0 & 0.355 & 0.79 & 0.05 &  23.5 & 0.354 & 1 & Yes & Yes \\
J0728$+$3835 & 1.25 & 0.85 &  67.6 & 0.392 & 1.25 & 0.01 & 170.6 & 0.393 & 4 & Yes & Yes \\
J0737$+$3216 & 1.00 & 0.67 &  98.8 & 0.239 & 0.97 & 0.10 &  97.8 & 0.233 & 2 & Yes & Yes \\
J0822$+$2652 & 1.17 & 0.88 &  68.2 & 0.370 & 1.14 & 0.01 &  10.5 & 0.365 & 2 & Yes & Yes \\
J0841$+$3824 & 1.41 & 0.79 &  91.4 & 0.242 & 1.36 & 0.05 &  10.2 & 0.236 & 2 & No & Yes \\
J0903$+$4116 & 1.29 & 0.90 & 161.3 & 0.396 & 1.27 & 0.02 & 142.4 & 0.393 & 2 & Yes & No \\
J0912$+$0029 & 1.63 & 0.56 &   8.2 & 0.288 & 1.62 & 0.10 &   5.1 & 0.286 & 1 & Yes & Yes \\
J0935$-$0003 & 0.87 & 0.69 &  22.2 & 0.160 & 0.81 & 0.13 &  27.0 & 0.152 & 1 & No & Yes \\
J0936$+$0913 & 1.09 & 0.89 & 160.1 & 0.315 & 1.09 & 0.02 &  16.7 & 0.315 & 2 & Yes & Yes \\
J0946$+$1006 & 1.38 & 0.81 & 159.2 & 0.355 & 1.39 & 0.08 & 157.9 & 0.357 & 2 & Yes & Yes \\
J0955$+$0101 & 0.91 & 0.82 &  62.5 & 0.458 & 1.03 & 0.27 &  27.6 & 0.499 & 2 & No & Yes \\
J0956$+$5100 & 1.33 & 0.63 & 146.2 & 0.356 & 1.30 & 0.11 & 144.2 & 0.351 & 1 & Yes & Yes \\
J0959$+$4416 & 0.96 & 0.92 &  57.4 & 0.310 & 0.96 & 0.00 &  35.0 & 0.310 & 2 & No & Yes \\
J0959$+$0410 & 0.99 & 0.86 &  66.9 & 0.397 & 1.01 & 0.07 & 142.1 & 0.402 & 2 & No & Yes \\
J1016$+$3859 & 1.09 & 0.78 &  46.4 & 0.414 & 1.06 & 0.08 &  38.9 & 0.406 & 2 & No & Yes \\
J1020$+$1122 & 1.20 & 0.80 & 135.8 & 0.413 & 1.21 & 0.10 & 152.6 & 0.416 & 2 & No & Yes \\
J1023$+$4230 & 1.41 & 0.87 & 170.4 & 0.435 & 1.40 & 0.03 & 168.8 & 0.433 & 3 & Yes & Yes \\
J1029$+$0420 & 1.01 & 0.84 &  93.9 & 0.378 & 1.10 & 0.17 &  48.0 & 0.401 & 1 & No & Yes \\
\hline
\end{tabular}
\tablecomments{Einstein radii $b_{\mathrm{SIE}}$ and $b_{\mathrm{LTM}}$
are quoted for an intermediate-axis normalization.
Mass minor-to-major axis ratios of SIE models are given by $q_{\mathrm{SIE}}$.
External shear values for LTM models are given by $\gamma_{\mathrm{ext}}$.
Position angles $PA$ (of SIE major axis) and $PA_{\gamma}$
(of LTM external shear) are measured in degrees E of N\@.
$L_{\mathrm{Ein, SIE}} / L_{\mathrm{deV}}$ and
$L_{\mathrm{Ein, LTM}} / L_{\mathrm{deV}}$ give luminosity
enclosed within SIE and LTM Einstein radii, evaluated using
B-spline luminosity models, as a fraction of de Vaucouleurs
total model luminosity.  $N_{\mathrm{src}}$ gives number
of source-plane components used to model background galaxy.
``Ring Subset?'' column indicates whether lens is included
in the subset of systems with full or partial Einstein-ring
lensed images.  ``Good $\sigma_{\mathrm{SDSS}}$?''
column indicates whether velocity dispersion is well-measured
in SDSS data.
}
\end{center}
\end{table}

\addtocounter{table}{-1}

\begin{table}[h]
\begin{center}
\caption{\label{dummy_3} (continued)}
\begin{tabular}{cccccccccccc}
\hline \hline
System Name & $b_{\mathrm{SIE}}$ & $q$ & $PA$ & $L_{\mathrm{Ein,SIE}}$ & $b_{\mathrm{LTM}}$ & $\gamma_{\mathrm{ext}}$ & $PA_{\gamma}$ & $L_{\mathrm{Ein,LTM}}$ & ~ & Ring & Good \\
(SDSS\ldots) & ($\arcsec$) & (SIE) & (SIE) & $ / L_{\mathrm{deV}}$ & ($\arcsec$) & (LTM) & (LTM) & $/ L_{\mathrm{deV}}$ & $N_{\mathrm{src}}$ & Subset? & $\sigma_{\mathrm{SDSS}}$? \\
\hline
J1032$+$5322 & 1.03 & 0.76 & 139.7 & 0.582 & 1.12 & 0.08 &  46.2 & 0.606 & 3 & No & Yes \\
J1100$+$5329 & 1.52 & 0.53 & 105.3 & 0.384 & 1.43 & 0.19 & 113.4 & 0.369 & 2 & No & No \\
J1103$+$5322 & 1.02 & 0.52 &  51.7 & 0.342 & 1.04 & 0.05 &  71.9 & 0.348 & 1 & Yes & Yes \\
J1106$+$5228 & 1.23 & 0.76 &  56.3 & 0.407 & 1.23 & 0.02 &  52.3 & 0.406 & 1 & Yes & Yes \\
J1112$+$0826 & 1.49 & 0.75 & 146.5 & 0.503 & 1.37 & 0.03 & 166.7 & 0.482 & 2 & No & Yes \\
J1134$+$6027 & 1.10 & 0.77 & 102.1 & 0.343 & 0.88 & 0.23 &  90.2 & 0.298 & 1 & No & Yes \\
J1142$+$1001 & 0.98 & 0.83 &  99.5 & 0.320 & 0.92 & 0.06 &  89.8 & 0.307 & 1 & No & Yes \\
J1143$-$0144 & 1.68 & 0.75 & 120.1 & 0.267 & 1.66 & 0.04 & 119.4 & 0.265 & 3 & No & Yes \\
J1153$+$4612 & 1.05 & 0.77 &  21.6 & 0.460 & 1.05 & 0.09 &  31.1 & 0.462 & 1 & Yes & Yes \\
J1204$+$0358 & 1.31 & 0.84 &  65.4 & 0.455 & 1.27 & 0.08 &  64.6 & 0.446 & 2 & Yes & Yes \\
J1205$+$4910 & 1.22 & 0.70 & 156.6 & 0.302 & 1.20 & 0.06 & 158.3 & 0.299 & 1 & Yes & Yes \\
J1213$+$6708 & 1.42 & 0.83 &  14.5 & 0.297 & 1.38 & 0.02 & 164.6 & 0.292 & 1 & No & Yes \\
J1218$+$0830 & 1.45 & 0.75 &  51.5 & 0.300 & 1.44 & 0.03 &  54.9 & 0.299 & 1 & No & Yes \\
J1250$+$0523 & 1.13 & 0.96 & 130.8 & 0.366 & 1.11 & 0.01 & 140.5 & 0.362 & 5 & Yes & Yes \\
J1251$-$0208 & 0.84 & 0.67 &  33.9 & 0.218 & 0.85 & 0.07 & 156.5 & 0.221 & 2 & No & No \\
J1402$+$6321 & 1.35 & 0.83 &  64.4 & 0.316 & 1.36 & 0.02 &  34.4 & 0.317 & 2 & Yes & Yes \\
J1403$+$0006 & 0.83 & 0.81 & 140.8 & 0.354 & 0.83 & 0.05 & 169.4 & 0.354 & 4 & Yes & Yes \\
J1416$+$5136 & 1.37 & 0.94 &  71.4 & 0.483 & 1.36 & 0.04 &  96.7 & 0.482 & 3 & No & Yes \\
J1420$+$6019 & 1.04 & 0.67 & 111.3 & 0.329 & 1.07 & 0.01 & 108.7 & 0.335 & 2 & Yes & Yes \\
J1430$+$4105 & 1.52 & 0.68 & 111.7 & 0.355 & 1.46 & 0.10 & 110.3 & 0.344 & 6 & Yes & Yes \\
J1432$+$6317 & 1.26 & 0.96 & 130.4 & 0.153 & 1.25 & 0.01 & 152.0 & 0.151 & 2 & No & Yes \\
J1436$-$0000 & 1.12 & 0.72 & 156.2 & 0.315 & 1.08 & 0.07 & 162.6 & 0.308 & 1 & No & Yes \\
J1443$+$0304 & 0.81 & 0.73 &  78.1 & 0.438 & 0.78 & 0.08 &  97.9 & 0.427 & 1 & No & Yes \\
J1451$-$0239 & 1.04 & 0.97 & 106.3 & 0.277 & 1.03 & 0.02 & 113.8 & 0.274 & 1 & No & Yes \\
J1525$+$3327 & 1.31 & 0.51 & 134.3 & 0.292 & 1.30 & 0.11 & 132.5 & 0.291 & 1 & No & Yes \\
J1531$-$0105 & 1.71 & 0.77 & 142.9 & 0.393 & 1.71 & 0.03 & 139.4 & 0.393 & 2 & Yes & Yes \\
J1538$+$5817 & 1.00 & 0.89 & 152.1 & 0.365 & 0.99 & 0.01 & 146.6 & 0.363 & 2 & Yes & Yes \\
J1621$+$3931 & 1.29 & 0.77 & 148.7 & 0.358 & 1.29 & 0.03 & 161.9 & 0.358 & 1 & No & Yes \\
J1627$-$0053 & 1.23 & 0.91 &  10.5 & 0.360 & 1.22 & 0.00 &  60.6 & 0.359 & 1 & Yes & Yes \\
J1630$+$4520 & 1.78 & 0.87 &  74.9 & 0.475 & 1.78 & 0.02 &  84.1 & 0.475 & 4 & Yes & Yes \\
J1636$+$4707 & 1.09 & 0.79 &  98.2 & 0.380 & 1.08 & 0.04 &  91.9 & 0.380 & 2 & Yes & Yes \\
J2238$-$0754 & 1.27 & 0.85 & 137.4 & 0.335 & 1.28 & 0.00 &  72.2 & 0.335 & 2 & Yes & Yes \\
J2300$+$0022 & 1.24 & 0.71 &  87.8 & 0.391 & 1.21 & 0.08 &  90.0 & 0.386 & 1 & Yes & Yes \\
J2303$+$1422 & 1.62 & 0.61 &  35.3 & 0.318 & 1.60 & 0.07 &  33.8 & 0.316 & 2 & Yes & Yes \\
J2321$-$0939 & 1.60 & 0.86 & 135.2 & 0.258 & 1.60 & 0.01 & 172.6 & 0.258 & 2 & Yes & Yes \\
J2341$+$0000 & 1.44 & 0.76 &  96.6 & 0.295 & 1.47 & 0.07 & 143.3 & 0.299 & 4 & Yes & Yes \\
\hline
\end{tabular}
\end{center}
\end{table}

\begin{table*}[h]
\begin{center}
\scriptsize
\caption{\label{specialnotes} Summary of unmodeled grade-A lenses}
\begin{tabular}{ll}
\hline \hline
System Name & Comments \\
\hline
SDSSJ0808$+$4706 & Nearby companion prevents simple SIE modeling. \\
SDSSJ1250$-$0135 & Complicated by spiral structure
  and asymmetric bulge in foreground galaxy. \\
SDSSJ1259$+$6134 & Faint \textsl{HST} and IFU features consistent with lensing;
  difficult to reconcile F814W and F435W images. \\
SDSSJ1618$+$4353 & Double foreground galaxy. \\
SDSSJ1718$+$6424 & Double foreground galaxy. \\
SDSSJ2141$-$0001 & Spiral/dust structure in foreground galaxy prevents
  acceptable model subtraction. \\
SDSSJ2302$-$0840 & Clear lens in IFU data; HST imaging inconclusive. \\
\hline
\end{tabular}
\end{center}
\end{table*}

\begin{figure*}[h]
\epsscale{1}
\centerline{\plotone{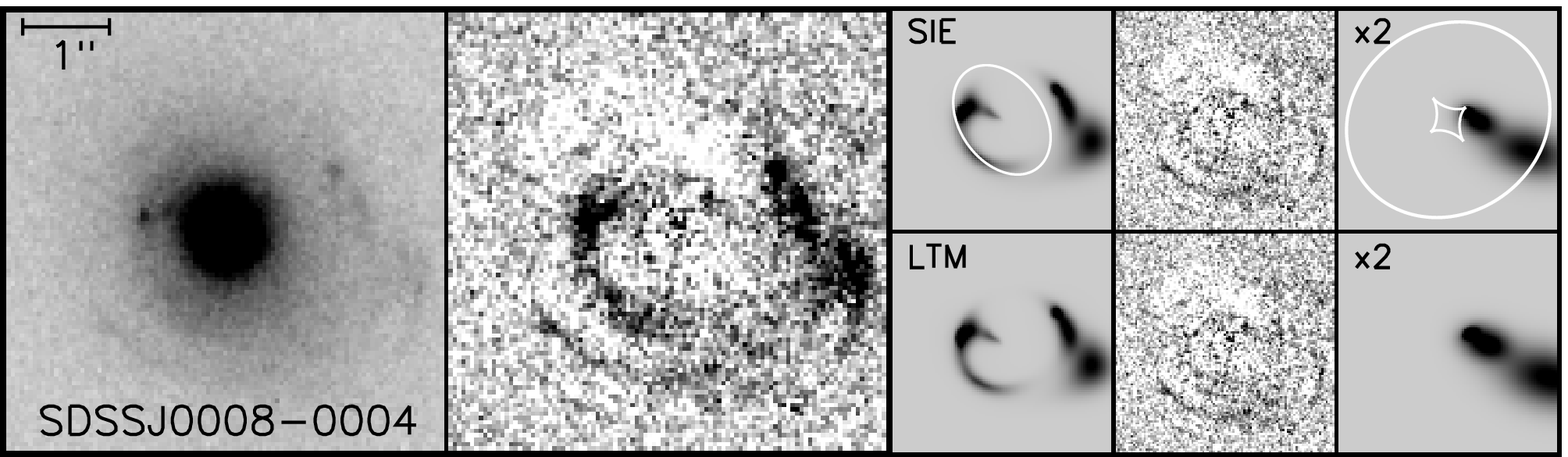}}
\centerline{\plotone{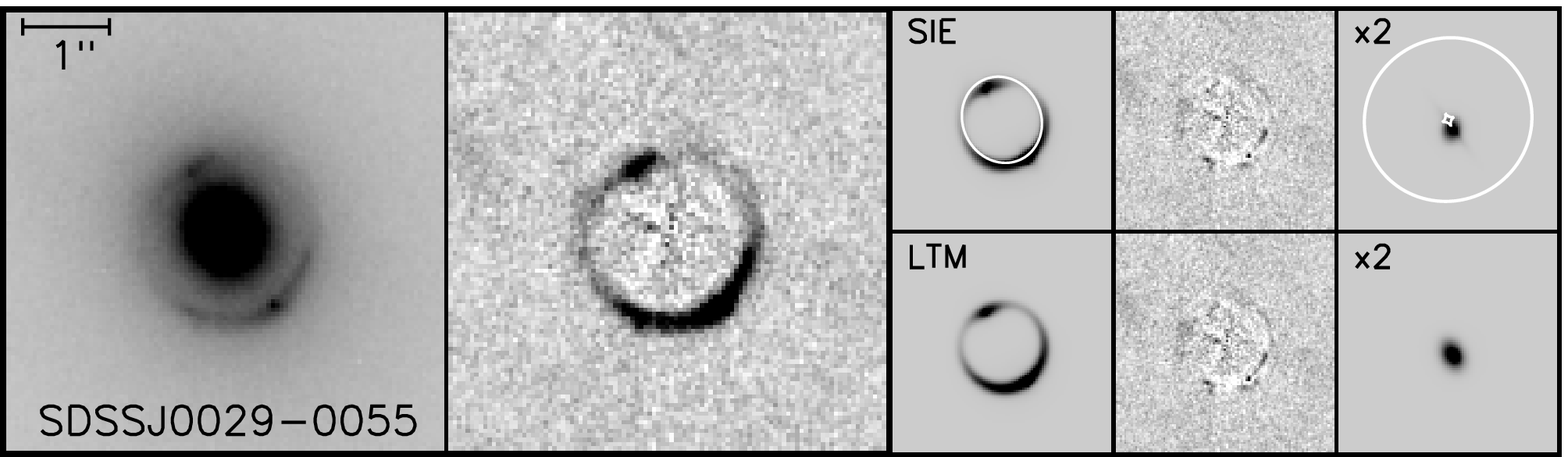}}
\centerline{\plotone{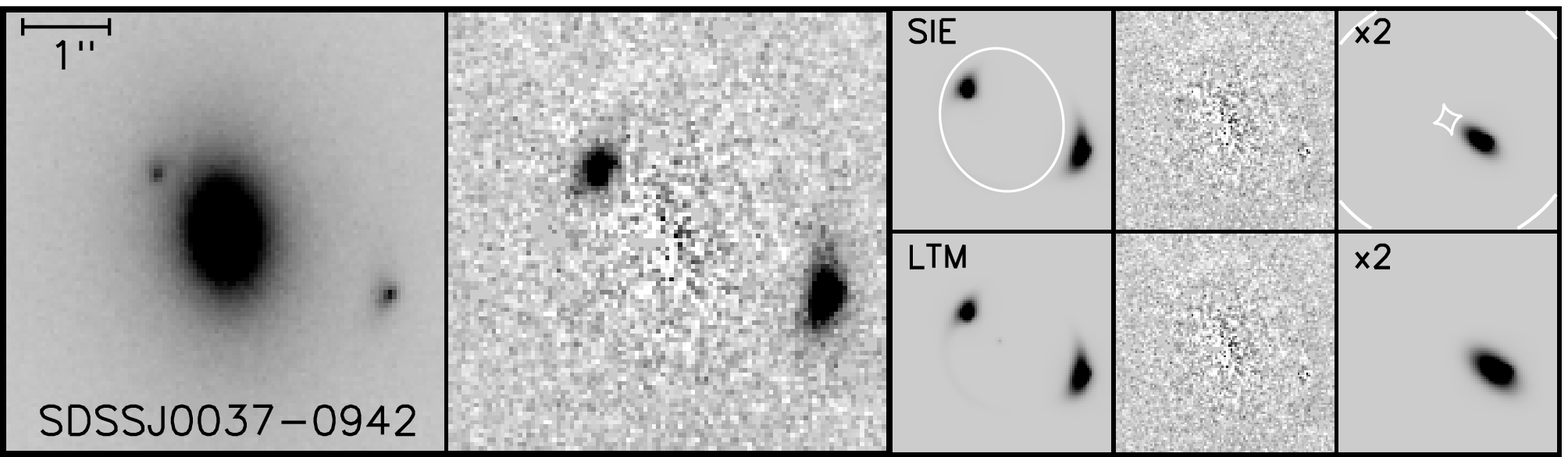}}
\caption{\label{prettypics} 
Lens models for grade-A SLACS \textsl{HST}-ACS strong
gravitational lens systems.
Leftmost large panels show direct F814W images, $5\arcsec \times 5\arcsec$ to a side,
with North up and East left.
Next large panels show same images, with B-spline model of foreground galaxy
subtracted, showing lensed features.  \textit{Top rows of smaller panels:}
Left: model prediction of best-fit SIE strong lens model for
features in residual data image, with critical curve in white;
Center: ``double-residual'' image,
after subtraction of B-spline and SIE models; Right: un-lensed source-plane
for best-fit SIE lens model, evaluated over a $2.5\arcsec \times 2.5\arcsec$
region and convolved with a $2\times$ de-magnified \textsl{HST} PSF for display purposes,
with caustics shown in white.  \textsl{Bottom rows of smaller panels}:
Same as top row, but for best-fit light-traces-mass (LTM) lens models and without
critical curves or caustics.
Grayscaling is linear in all images, ranging from $-0.25 X$ (white) to $X$ (black).
For the direct images, $X$ is set to the 97th-percentile image value as determined
from the smooth B-spline model.  For the residual and lens-model images, $X$ is
set to the 99th-percentile image value as determined from the SIE lens-model image.
Figures for all 63 lens models are available through the electronic version
of the \textsl{Astrophysical Journal}, or through the website of the first author.}
\end{figure*}

\end{document}